\documentclass{aa}

\usepackage{graphicx}
\usepackage{txfonts}
\usepackage{xcolor}

\begin{document}

   \title{Asteroseismology of the multiperiodic field SX Phe pulsator\\ BL~Camelopardalis}

\author{
W. Szewczuk\inst{1}
\and
J. Daszy\'nska-Daszkiewicz\inst{1}
\and
P. Walczak\inst{1}
\and
E. Rodr\'iguez\inst{2}
}

\institute{
University of Wroc{\l}aw, Faculty of Physics and Astronomy, Astronomical Institute, Kopernika 11, PL-51-622 Wroc{\l}aw, Poland\\
\email{wojciech.szewczuk@uwr.edu.pl}
\and
Instituto de Astrof\'isica de Andaluc\'ia, CSIC, PO Box 3004, E-18080 Granada, Spain
}

   \date{Received September 15, 1996; accepted March 16, 1997}

  \abstract
   {BL~Camelopardalis (BL~Cam) is the most metal-poor SX Phoenicis star known in the Galactic field, making it an excellent test bed for studies of Population II stellar structure and evolution, as well
   as for investigating the formation channels of blue straggler stars.}
   {We aim to constrain fundamental stellar parameters and pulsational properties of BL~Cam and identify the nature of its observed oscillation modes.}
   {We analysed high-precision space-based data from the TESS mission and long-term ground-based observations from the Zwicky Transient Facility (ZTF) survey. The ZTF data were primarily used to derive the orbital parameters of the system, while the TESS light curves enabled us to extract a rich oscillation spectrum. We identified multiple pulsation frequencies and performed seismic
   modelling using a Bayesian approach.}
   {We detected numerous pulsation frequencies, including a dominant radial mode and additional non-radial components.
   We predicted all fitted modes in our seismic models as excited. The models reproduced the two highest amplitude-independent frequencies (i.e. the radial fundamental mode and
   a dipole mode) and suggested the possible presence of additional radial overtones.
   The inferred stellar parameters confirm the extremely low metallicity of BL~Cam. We also confirm its binary nature, with an orbital period of 144 days.}
   {BL~Cam provides a valuable benchmark for testing stellar models of metal-poor pulsators. The combination of space-based photometry and Bayesian seismic modelling enables robust constraints to be placed on its internal structure and pulsation properties.}

   \keywords{
   asteroseismology -- Stars: evolution -- Stars: binaries -- Stars: oscillations -- Stars: Population II -- Stars: individual: BL~Cam
               }

   \maketitle
    \nolinenumbers

\section{Introduction}

SX Phoenicis stars are short-period pulsating variables whose properties are similar to those of $\delta$ Scuti stars.
However, unlike $\delta$ Sct stars, which belong to Population I, SX Phe stars are members of Population II.
BL~Camelopardalis (BL Cam) is one of the few SX Phoenicis stars found outside stellar clusters.
The paucity of field SX Phe stars is due to the difficulty in identifying them.
In the case of stellar clusters, these pulsating variables are located on the colour-magnitude diagram (CMD)
in a well separated region of blue stragglers \citep{Sandage1953}. In other words, they are hotter and brighter than the cluster stars at the main sequence turn-off.
The field SX Phe stars are identified as having low metallicity coupled with the high spatial velocities and space distributions typical of old and halo stars \citep[e.g.][]{Preston2000}. Most field SX Phe stars have been discovered to exist within binary (multiple) systems, including BL~Cam.

Here, we present the first seismic analysis of the main component of the BL~Cam system.
In Sect.\,\ref{sec:2}, we provide details about the system and pulsations of the primary.
Sect.\,\ref{sec:3} is devoted to the new binary solution derived from Zwicky Transient Facility (ZTF) data.
The results of frequency analysis of the TESS light curve are given in Sect.\,\ref{sec:4}.
Sect.\,\ref{sec:5} is devoted to the mode identification of the dominant frequency as well as the secondary independent frequency.
In Sect.\,\ref{sec:6}, the results of seismic modelling are presented.
The summary in Sect.\,\ref{sec:7} concludes the paper.

\section{Basic parameters of BL~Cam}
\label{sec:2}
BL~Cam was mentioned in the literature for the first time by \citet{Giclas1970}
as a possible white dwarf candidate, designated as GD 428.
It was discovered to be a variable star with a short period of about 0.039\,d
and a mean brightness of $V\approx 13$ mag by \citet{Berg1977ApJ...215L..25B}.
\cite{1985PASP...97..715M} reported an  upper limit of the projected rotational velocity of $V \sin i = 18\,\mathrm{km\,s^{-1}}$.

BL~Cam is a unique variable because of its lowest metallicity amongst field SX Phe stars.
Its spectrum displays mostly hydrogen lines \citep{McNamara1978PASP...90..275M, 2011RMxAC..40..266A}, while some determinations of its metallicity indicate that the
[m/H] value might be even below $-2.0$, which translates to a metallicity by mass of $Z\lesssim 0.0002$.
The following values can be found in the literature: $\mathrm{[m/H]}= -1.5 \pm 0.2$ \citep{1983A&A...121..250A, 1991A&A...247...77R},
$\mathrm{[m/H]} = -2.4 \pm 0.2$ \citep{1994AJ....108..222N, 1997PASP..109.1221M},
$\mathrm{[m/H]}= -2.2 \pm 0.3$  \citep[Gaia,][]{2023A&A...674A...1G}, and
$\mathrm{[m/H]}= -1.2 \pm 0.3$ \citep{2021RMxAA..57..419P}.

The total range of the effective temperature found in the literature is $T_{\rm eff}\in(7043, 8153)$\,K
\citep{McNamara1978PASP...90..275M, 1983A&A...121..250A,1997PASP..109.1221M, 2021RMxAA..57..419P, 2022Ap.....65..456A, 2023A&A...674A...1G}.
The distance to BL~Cam is $d = 669^{+16}_{-8}$\,pc, as derived by \cite{2022A&A...658A..91A} using the Gaia DR3 parallax and StarHorse2 model. Combining this distance with the apparent  magnitude of $V=12.930\pm 0.018$ \citep{Berg1977ApJ...215L..25B}, along with a colour excess of
$E(B-V)=0.350\pm0.014$ \citep[Bayestar2019 extinction maps,][]{2019ApJ...887...93G} and bolometric correction from \cite{1996ApJ...469..355F}
(see also \citealt{2010AJ....140.1158T} for corrected coefficients), we were able to determine a luminosity of $\log L/L_{\sun}= 0.65 \pm 0.16$.

Since BL~Cam is a binary system with an uncertain contribution from the companion, we computed the luminosity both under
the assumption that the observed flux originates from a single star and that it is equally shared between two identical stars.
The quoted value represents the mean of these two cases, with the uncertainty spanning the full plausible range.

\begin{figure}
	\includegraphics[width=\columnwidth,clip]{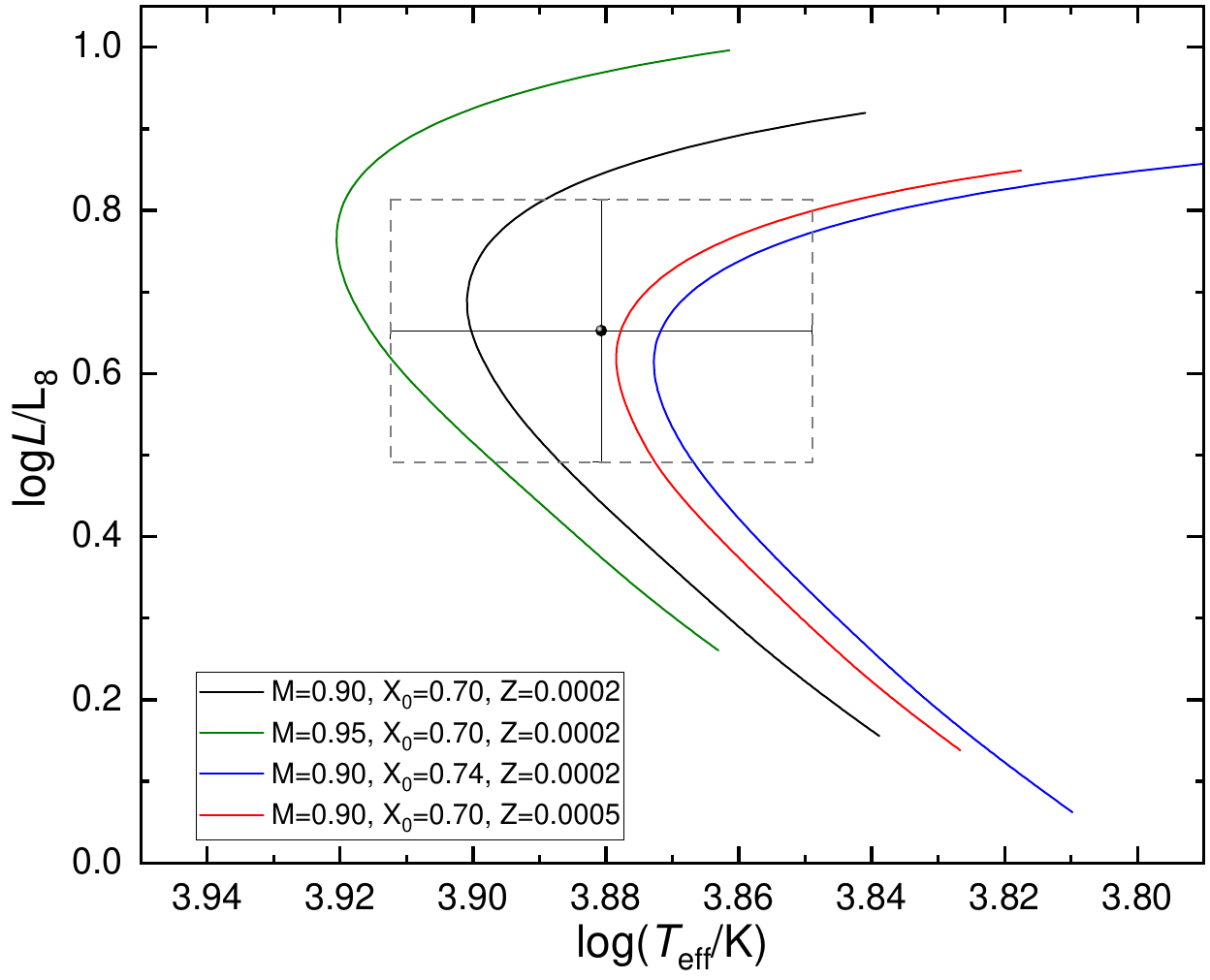}
	\caption{Position of BL~Cam on the HRD. Evolutionary tracks were computed with the OPAL opacity tables, the solar mixture from AGSS09, the MLT parameter $\alpha_{\rm MLT}=0.5$ and an initial rotation of $V_{{\rm rot},0}=20\,\mathrm{km\,s^{-1}}$. The effects of mass, $M$, metallicity, $Z$, and the initial hydrogen abundance, $X_0$, are also shown.
			}
\label{fig_HR}
\end{figure}

The position of BL~Cam on the Hertzsprung-Russell diagram (HRD) is shown in Fig.\,\ref{fig_HR}.
Four evolutionary tracks are plotted for guidance, displaying the effect of mass, initial hydrogen abundance, $X_0$, and metallicity, $Z$.
All these evolutionary models and the stellar models discussed throughout this paper
were computed with the Warsaw-New Jersey code using the OPAL opacity tables \citep{Iglesias1996}
and the solar chemical mixture from \citet{Asplund2009}, hereafter referred to as AGSS09.
The code takes into account the mean effect of the centrifugal force. Rigid rotation and conservation of global angular momentum during evolution were assumed in this work.
Convection in the stellar envelope was treated within the framework of standard mixing-length theory (MLT). The OPAL2005 equation of state was adopted \citep{Rogers1996,Rogers2002}.

BL~Cam has been intensively studied photometrically in recent decades
\citep[e.g.][]{Berg1977ApJ...215L..25B, 1997PASP..109...15H, 1999MNRAS.308..631Z, 2006A&A...451..999F,  2007A&A...471..255R, 2019PASP..131f4202Z,  2022Ap.....65..456A}.
The  multisite campaign by \cite{2007A&A...471..255R} revealed an exceptionally rich pulsational content with 25 significant peaks
of which 23 correspond to independent modes.
This makes BL~Cam one of the most multiperiodic SX Phoenicis stars known to date, exhibiting variable amplitudes and complex mode behaviour.

O-C analyses have long suggested that BL~Cam is a likely binary or even a hierarchical multiple system.
However, different authors have obtained markedly different orbital solutions.
\cite{2006A&A...451..999F} found that the observed variations of the main period can be described by a secular increase at a rate of
$dP/Pdt=117(\pm 3)\times 10^{-9}\, \mathrm{yr}^{-1}$,
combined with a perturbation from a companion star on an eccentric orbit $(e=0.7\pm 0.2)$ with
a period of $P_\mathrm{orb}=10.5\pm0.2$\,yr.
\cite{2008ChJAA...8..237F}
derived $e=0.7 \pm 0.2$  and $P_\mathrm{orb}=9.7 \pm1.6$\,yr.
\cite{2010A&A...515A..39F} reported $e=0.19\pm0.01$
and $P_\mathrm{orb}=144.19 \pm 0.02$ days for one component, along with evidence of a third body
on an orbit with
$e=0.7$ and $P_\mathrm{orb}=9.28 \pm 0.07$\,yr.
\cite{2019PASP..131f4202Z} obtained
$e=0.80\pm0.07$
and $P_\mathrm{orb}=14.01 \pm 0.09$\,yr,
while \cite{2021RMxAA..57..419P} proposed yet another solution with
$e=0.15$ and $P_\mathrm{orb}=55.2$\,yr.

\section{ZTF photometry and the binary solution}
\label{sec:3}

\begin{figure}
	\includegraphics[width=\columnwidth,clip]{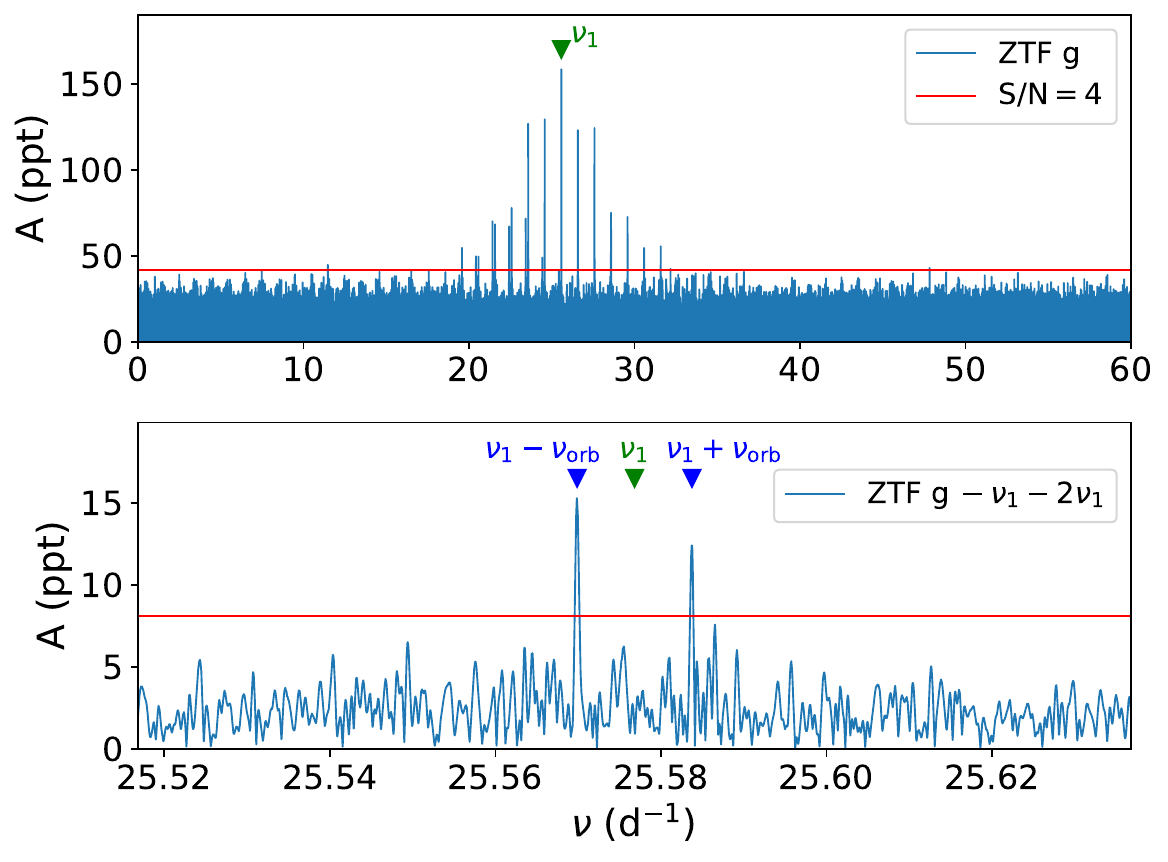}
	\caption{LS periodograms of the ZTF $g$-band light curve of BL~Cam. The top panel shows the periodogram of the original data plotted over
     $0-60\,\mathrm{d}^{-1}$. The bottom panel presents the periodogram after the pre-whitening at $\nu_1$ and $2\nu_1$, zoomed in on the vicinity of $\nu_1$.
     Significant frequency peaks are marked with triangles: the green symbol indicates $\nu_1$, while the blue triangles show the sidelobes induced
     by orbital motion. In the lower panel, the position of $\nu_1$ is indicated for clarity. The red lines mark our adopted significance threshold.}
\label{fig:ZTF_period}
\end{figure}

ZTF is a wide-field time-domain survey that has been operating at the Palomar 48-inch Samuel Oschin Schmidt telescope since 2018 \citep{2019PASP..131a8002B}. Equipped with a 47\,deg$^2$ field-of-view camera, ZTF conducts high-cadence photometric monitoring of the northern sky in the $g$, $r$, and $i$ bands. In this work, we used the publicly available ZTF light curves of BL~Cam obtained in the $g$ and $r$  filters.
After discarding all data points with non-zero catflags as unreliable, the
$g$-band light curve contained 418 measurements spanning 2297 days (from 13 July 2018 to 26 October 2024),
corresponding to a mean sampling of one observation every 5.5 days and a Rayleigh frequency resolution of $\Delta \nu_R(g)=0.000435$\,d$^{-1}$.
In the $r$ band, 529 reliable measurements covered 2395 days, yielding a mean cadence of one point every 4.5 days and $\Delta \nu_R(r)=0.000418$\,d$^{-1}$.

For both filters, we applied our standard frequency analysis procedure,
computing periodograms and performing successive pre-whitening \citep[see e.g.][]{2021MNRAS.503.5894S, 2024MNRAS.532.1140D}.
We adopted a signal-to-noise ratio (S/N) of 4 as the significance threshold \citep{1993A&A...271..482B, 1997A&A...328..544K}.
A representative Lomb-Scargle (LS) periodogram of the ZTF $g$ band light curve is shown in Fig.\,\ref{fig:ZTF_period}.
The top panel displays the periodogram of the original data plotted over the frequency range $0-60\,\mathrm{d}^{-1}$, where the dominant signal at
$\nu_1=25.57681\,\mathrm{d}^{-1}$ is clearly visible. The bottom panel shows the periodogram after pre-whitening at
$\nu_1$ and $2\nu_1$, zoomed in on the vicinity of $\nu_1$.
The parameters of our variability model derived from these data are listed in Table\,\ref{tab:freq_ZTF}.

\begin{table}
\small
\caption{Frequency analysis results for BL~Cam from ZTF $g$- and $r$-band photometry.}
\label{tab:freq_ZTF}
\centering
\begin{tabular}{cccccc}
\hline
ID & $\nu$ (d$^{-1}$) & $A$ (ppt) & $\phi$ (0-1) & S/N & Remarks\\
\hline
\multicolumn{6}{c}{$g$ filter} \\
\hline
1  & 25.576817(2) & 157(1)  & 0.609(2) & 15.2 & \\
   & 51.15363(1)  & 37(1)   & 0.030(9) & 12.2 & $2\nu_1$\\
   & 25.56988(3)  & 15(1) & 0.53(2) & 7.5  & $\nu_1-\nu_\mathrm{orb}$\\
   & 25.58375(3)  & 13(1) & 0.16(3) & 6.6  & $\nu_1+\nu_\mathrm{orb}$\\
   & 51.14668(4)  & 9(1)  & 0.96(3) & 5.5  & $2\nu_1+\nu_\mathrm{orb}$\\
2  & 25.25222(5)  & 8(1)  & 0.29(4) & 4.9  & \\
   & 76.73046(6)  & 6(1)  & 0.48(6) & 4.1  & $3\nu_1$\\
3  & 84.07345(7)  & 5(1)  & 0.73(7) & 4.1  & \\
4  & 27.10710(8)  & 5(1)  & 0.89(7) & 4.0  & \\
5  & 115.83085(9) & 4(1)  & 0.73(8) & 4.1  & \\
\hline
\multicolumn{6}{c}{$r$ filter} \\
\hline
1 & 25.576816(3) & 111.5(8) & 0.608(2) & 17.0 & \\
  & 51.15362(1)  & 26.1(8)  & 0.05(1)  & 12.6 &  $2\nu_1$\\
  & 25.56988(2)  & 12.1(8)  & 0.54(2)  & 8.1  & $\nu_1-\nu_\mathrm{orb}$\\
  & 25.58377(4)  & 7.6(8)   & 0.14(4)  & 6.5  & $\nu_1+\nu_\mathrm{orb}$\\
  & 76.73043(5)  & 6.0(8)   & 0.54(5)  & 4.9  & $3\nu_1$\\
2 & 25.25212(5)  & 6.0(8)   & 0.43(4)  & 4.9  & \\
  & 51.14682(5)  & 5.8(8)   & 0.83(5)  & 5.0  & $2\nu_1-\nu_\mathrm{orb}$\\
3 & 11.48079(6)  & 4.8(8)   & 0.26(6)  & 4.3  & \\
4 & 108.68046(7) & 4.0(8)   & 0.56(7)  & 4.2  & \\
\hline
\end{tabular}
\tablefoot{Only the independent frequencies are numbered in the first column. The phases ($\phi$) are given for $t = \mathrm{BJD} - 2458216$\,d.
Amplitudes are expressed in parts per thousand (ppt).}
\end{table}

\begin{table}
\small
\caption{Orbital parameters of BL~Cam derived from the frequency modulation analysis.}
\label{tab:orb_sol_ZTF}
\centering
\renewcommand{\arraystretch}{1.4}
\begin{tabular}{lcc}
\hline
Parameter & $r$ filter & $g$ filter\\
\hline
$P_{\rm orb}$ (d)                  & $144.01^{+0.48}_{-0.38}$  & $144.20^{+0.40}_{-0.39}$ \\
$a\sin i$ (AU)     &              $0.194^{+0.011}_{-0.011}$    &  $0.196^{+0.010}_{-0.010}$ \\
$f(m_1,\,m_2,\,\sin i)^{a}~(M_{\sun})$    & $0.047^{+0.009}_{-0.008}$ & $0.049^{+0.008}_{-0.007}$\\
$f(m_1,\,m_2,\,\sin i)^{b}~(M_{\sun})$    & $0.047^{+0.008}_{-0.008}$ & $0.049^{+0.008}_{-0.007}$\\
$e^b$                                & $0.20^{+0.10}_{-0.09}$   & $0.21^{+0.09}_{-0.08}$ \\
\hline
\end{tabular}
\tablefoot{$^{a}$Derived using the two-sidelobe fit. $^{b}$Derived using the four-sidelobe fit.}
\end{table}

In both filters we clearly detected the well-known dominant frequency of
$\nu_1=25.57681\,\mathrm{d}^{-1}$
and its first two harmonics, as well as the secondary peak at
$\nu_2=25.252\,\mathrm{d}^{-1}$ \citep[see e.g.][]{2007A&A...471..255R}.
The other frequencies listed as independent are most likely spurious,
since they differ between the two filters and are absent in the TESS light curves (see Sect.\,\ref{sec:4}).
One aspect of particular interest is the presence of symmetrically spaced side peaks around $\nu_1$,
forming an equidistant triplet whose frequency separation matches the orbital frequency reported by \citet{2010A&A...515A..39F}.
The triplet pattern in the $g$-band is illustrated in Fig.\,\ref{fig:ZTF_period}.
We interpreted these side peaks as arising from the light-time effect induced by orbital motion. This allowed us to derive an orbital
solution using the frequency modulation (FM) method \citep[e.g.][]{2012MNRAS.422..738S}.
Parameter uncertainties were estimated from Monte Carlo simulations,
taking into account the measurement errors on both amplitudes and frequencies.
The quoted uncertainties correspond to the 16th and 84th percentiles of the resulting distributions.
The obtained orbital parameters are given in Table\,\ref{tab:orb_sol_ZTF}.
Our orbital period and value of $a\sin i$ are consistent (within the errors) with those determined by \citet{2010A&A...515A..39F} from O-C analysis.
We were not able to compare the mass function, as it was not provided in that study.
In principle, determination of the orbital eccentricity within the FM framework requires the amplitudes of four sidelobes, separated from the central pulsation frequency
by $\pm \nu_\mathrm{orb}$ and  $\pm 2\nu_\mathrm{orb}$. In our data, only the inner pair at $\pm \nu_\mathrm{orb}$ was detected, while the outer sidelobes at
$\pm 2\nu_\mathrm{orb}$ remained below the detection threshold. We therefore labelled the mass function derived from the actual observations as the 'two-sidelobe fit'
in Table\,\ref{tab:orb_sol_ZTF}. For comparison, we also constructed a synthetic `four-sidelobe model' by adding the undetected
$\pm 2\nu_\mathrm{orb}$ components to the detected ones and fitting this full set simultaneously.
This yielded the same value of mass function and estimation of an eccentricity $e\approx0.2\pm0.1$ (see the four-sidelobe fit in Table\,\ref{tab:orb_sol_ZTF}).
Because the amplitudes of the undetected sidelobes are highly uncertain, the resulting eccentricity estimate carries a correspondingly large  error; however, it remains consistent with the value $e=0.19\pm0.01$ reported by \citet{2010A&A...515A..39F}.

\section{Pulsational frequencies from the TESS data}
\label{sec:4}

The Transiting Exoplanet Survey Satellite \citep[TESS,][]{2015JATIS...1a4003R} is a NASA space telescope launched in 2018 to obtain high-precision, wide-field photometry.
It observes the sky in consecutive $24^\circ \times 96^\circ$ sectors, typically providing nearly continuous coverage for about 27 days per sector.
Although it has primarily been designed to detect exoplanet transits, its high duty cycle and stable red-optical bandpass (approximately 600--1000 nm)
make it a powerful facility for stellar variability studies.
In addition to the 2-min (and, in the extended mission, 20-s) cadence data for selected targets \citep[SPOC,][]{2016SPIE.9913E..3EJ},
the Quick-Look Pipeline \citep[QLP,][]{2020RNAAS...4..204H, 2020RNAAS...4..206H} produces light curves from full-frame
images (FFI) for millions of stars. These have evolved from 30-min sampling in the primary mission to 10-min and, more recently, 200-s cadence in the extended mission.

BL~Cam was observed by TESS in Sectors 19, 59, and 86. For Sector 19 we used the publicly available SPOC light curve and adopted the PDCSAP fluxes.
For Sectors 59 and 86, we used the publicly available QLP light curves derived from the full-frame images (FFIs).
All light curves were obtained from the Mikulski Archive for Space Telescopes (MAST).

For each sector, only measurements with \texttt{QUALITY}=0 were retained. The light curve in each sector was subsequently normalised by dividing it by a linear fit to remove residual long-term trends. A visual inspection of the resulting light curves revealed no significant outliers.

The basic properties of each dataset are summarised in Table\,\ref{tab:tessdata}, including the time span, number of data points, cadence, and the corresponding Rayleigh frequency resolution.
Since the three sectors are separated by long gaps, we started by analysing each sector individually.

In the next step of our study, we adopted a more conservative significance criterion of $\mathrm{S/N}>5$ \citep{2021AcA....71..113B},
which is more appropriate for high-precision space-based photometry than the commonly used threshold of $\mathrm{S/N}>4$.
In contrast to our analysis of the ZTF data, where the noise level was estimated as the average amplitude over the entire LS periodogram,
here it was calculated locally as the mean amplitude within a $10\,\mathrm{d^{-1}}$ wide frequency window centred on the tested frequency.
As in the ZTF analysis, the periodograms were computed up to a maximum frequency of $120\,\mathrm{d^{-1}}$.

\begin{figure*}
	\includegraphics[width=2\columnwidth,clip]{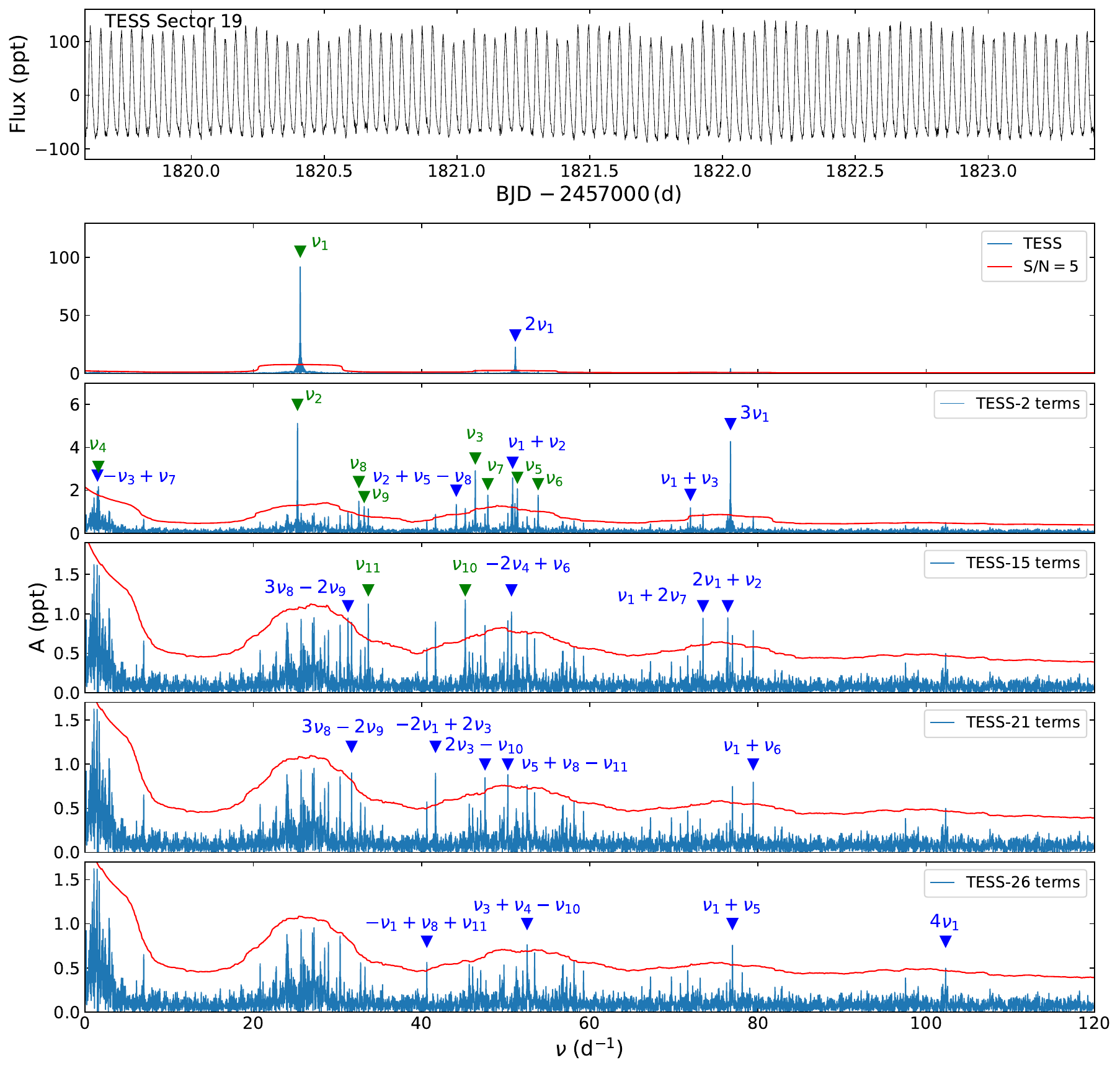}
	\caption{Top: Short segment of the TESS Sector 19 light curve of BL~Cam. Second panel: Amplitude LS periodogram calculated from the original data. Third, fourth, fifth, and bottom panels: Periodograms of the residuals after pre-whitening the 2, 15, 21, and 26 frequencies, respectively. Green and blue triangles denote independent and combination or harmonic frequencies. The red curve corresponds to the adopted $\mathrm{S/N}=5$ significance threshold for the corresponding stage of the pre-whitening procedure.}
\label{fig:periodograms_S19}
\end{figure*}

As an example, a short segment of the TESS Sector 19 light curve is shown in the top panel of Fig.\,\ref{fig:periodograms_S19}.
The subsequent panels present amplitude LS periodograms calculated from the original Sector 19 data (second panel from the top) and from the data pre-whitened by 2, 15, 21, and 26 frequencies, respectively.
After pre-whitening all 30 frequencies listed in Table\,\ref{tab:freq_TESS_s19}, no peak remains above our adopted significance threshold.

All significant frequencies are marked with triangles, where green and blue symbols denote independent and combination or harmonic frequencies, respectively.
The red curve shown in each panel corresponds to the adopted significance threshold of $\mathrm{S/N}=5$, calculated for the particular pre-whitening stage represented by that panel.
Consequently, the displayed threshold is directly applicable only to the highest amplitude peak identified at that stage.
Frequencies with lower amplitudes were generally detected during subsequent pre-whitening steps,
after the removal of additional signals had reduced the local noise level. Their actual S/Ns at the time of detection
might therefore be higher than suggested by the threshold shown in a given panel.

Following our standard procedure \citep[see, e.g.][]{2021MNRAS.503.5894S,2024MNRAS.532.1140D}, we searched for unresolved frequency pairs.
Frequencies separated by less than 2.5 times the Rayleigh resolution were considered unresolved following the conservative criterion of \cite{1978Ap&SS..56..285L}.
In such cases only the frequency with the larger amplitude was retained in the final solution.
This procedure identified only one unresolved frequency pair in the Sector 19 data, where the lower-amplitude component was excluded from the final fit.

All harmonics and combination frequencies were searched for within the Rayleigh resolution. The results are presented in Tables\,\ref{tab:freq_TESS_s19}--\ref{tab:freq_TESS_s86}.
As can be seen, the frequency content detected in different sectors, especially at low amplitudes, is not identical. The smallest number of significant frequencies was
found in Sector 86, which is most likely a consequence of the smallest number of available data points.

Our frequency set also differs from that reported by \citet{2007A&A...471..255R}.
However, the dominant modes and several additional frequencies are consistent between both studies, while the discrepancies mainly concern low-amplitude signals.
In particular, we did not detect their reported frequency of $f_6=32.64641$\,d$^{-1}$, which had an amplitude of 2.58\,mmag in the $V$ filter during the ground-based campaign (August~2005 to March~2006).
This could suggest that its amplitude decreased over the last two decades to a level below the TESS detection threshold.
This frequency was interpreted as a radial first-overtone mode based on the period ratio with the dominant frequency.
However, we detect another frequency $\nu_8 = 32.546$\,d$^{-1}$ (see Table\,\ref{tab:freq_TESS_s19}), which might also correspond to a radial mode, based on its period ratio with $\nu_1$. Moreover, this signal is present in the data of \citet{2007A&A...471..255R},
where it was marked as $f_{16}=32.54533$~d$^{-1}$.

The differences between the frequencies detected in individual TESS sectors, as well as between our results and those of \citet{2007A&A...471..255R}, might result from a range of plausible causes.
First, BL~Cam is known to exhibit significant amplitude variability \citep{2007A&A...471..255R}, which could lead to the appearance or disappearance of low-amplitude signals.
Second, the TESS observations provide only limited coverage of the binary orbital phase (see Appendix~\ref{app1}) for the $\sim$144\,d orbit,
which could introduce apparent frequency shifts due to the light-travel time effect.
A comparison of the frequencies obtained from individual sectors and with those reported by \citet{2007A&A...471..255R} is provided in Appendix~\ref{app1}.

On the other hand, the dominant frequencies in terms of amplitude appear to be relatively stable. The amplitude of $\nu_1$ is about 90\,ppt in all analysed TESS sectors,
indicating no significant variation over the time span of the observations. A similar behaviour is observed for $\nu_2$, which has an amplitude of approximately 5\,ppt.
We stress, however, that small differences in the amlplitudes between sectors may partly result from the use of different photometric apertures in the TESS data.

Since the central wavelength of the TESS passband is similar to that of the $I$ filter in the $UBVRI$ system,
we can compare our results with those of \citet{2007A&A...471..255R}. The TESS amplitudes of $\nu_1$ and $\nu_2$ expressed in magnitudes
are approximately 98\,mmag and 5.4\,mmag, respectively.
These values are comparable to the amplitudes reported in the $I$ band by \citet{2007A&A...471..255R}; namely, 91\,mmag for $\nu_1$ and 4.3\,mmag for $\nu_2$.
In the ZTF $g$-band, the amplitudes of $\nu_1$ and $\nu_2$ are about 170\,mmag and 9\,mmag, respectively. The central wavelength of the ZTF $g$ filter lies approximately between those of the $B$ and $V$ bands. For comparison, \citet{2007A&A...471..255R} reported amplitudes of 184 and 154\,mmag in the $B$ and $V$ bands, respectively, for $\nu_1$, and 8 and 7\,mmag for $\nu_2$.

Finally, we attempted to merge all available TESS observations and perform a frequency analysis of the combined dataset.
However, such an analysis may be affected by the large gaps between the sectors, as well as by the intrinsic amplitude variability and possible
systematic differences caused by the use of different photometric masks in different sectors. The list of frequencies detected in the whole
dataset is presented in Appendix~\ref{app1}.

Furthermore, we detected a small number of significant peaks in the low-frequency domain ($\sim$1.5~d$^{-1}$).
Similar low-frequency signals were already reported in the ground-based photometry of BL~Cam by \citet{2007A&A...471..255R}. These authors noted that a number of peaks appeared mainly below 2\,d$^{-1}$, although they considered them possibly spurious due to their strong dependence on the analysed data subset.
On the other hand, low-frequency variability in SX Phe stars may also be of intrinsic origin, as hybrid pulsators exhibiting both $p$- and $g$-mode oscillations
have been reported in the literature \citep[e.g.][]{2018AcA....68..237R}.
However, the low-frequency region is particularly sensitive to residual instrumental trends and detrending procedures, and thus the exact frequencies may depend on the analysed data subset.
In addition, individual TESS sectors sample different parts of the $\sim$144 d orbital cycle
and the light-travel time effect could introduce small apparent frequency shifts between sectors.
When the sectors are combined, the large temporal gaps and incomplete orbital-phase
coverage may further distort the low-frequency spectrum and produce additional spurious peaks.

Nevertheless, one low-frequency signal appears to be securely detected. This is a peak near 1.46\,d$^{-1}$ detected in the combined dataset
($\nu=1.46282$\,d$^{-1}$) and in individual sectors, e.g. $\nu\simeq1.450$\,d$^{-1}$ in Sector 19 and $\nu \simeq 1.460$\,d$^{-1}$ in Sector 59.
Given the limited Rayleigh resolution of single sectors, these values are consistent within the uncertainties and can be treated as the same feature.
The signal is not detected in Sector 86, most likely due to the smaller number of data points and the reduced sensitivity,
as reflected by the overall smaller number of extracted frequencies in that sector.
Therefore, although the presence of a coherent low-frequency variability component cannot be excluded,
we take the low-frequency content with caution.

\begin{table}
\small
\caption{Summary of TESS photometry for BL~Cam.}
\label{tab:tessdata}
\centering
\begin{tabular}{lcccc}
\hline
Sector & Time span (BJD) & $N_{\rm points}$ & Cadence (s) & $\Delta \nu_R$ (d$^{-1}$) \\
\hline
19  & $2458713 - 2458739$ & 16744 & 120     & 0.040 \\
59  & $2459439 - 2459465$ & 9804 & 200     & 0.040 \\
86  & $2460123 - 2460149$ & 5552 & 200     & 0.038 \\
\hline
\end{tabular}
\tablefoot{For SPOC, we used the PDCSAP fluxes, while QLP provides a single light curve extracted from FFIs.}
\end{table}

\begin{table}
\small
\caption{Frequency analysis results for BL~Cam from TESS photometry (Sector 19).}
\label{tab:freq_TESS_s19}
\centering
\begin{tabular}{ccccc}
\hline
ID & $\nu$ (d$^{-1}$) & $A$ (ppt)  & S/N & Remarks\\
\hline
   1  & 25.57734(3)  &     91.83(9)          & 58.6 &  \\
      & 51.1554(1)   &     22.96(9)          & 44.9 & $2\nu_1$  \\
   2  & 25.2530(6)   &      5.15(9)          & 19.9 &   \\
      & 76.7335(7)   &      4.34(9)          & 25.2 & $3\nu_1$  \\
   3  & 46.392(1)    &      2.92(9)          & 13.1 &   \\
      & 50.829(1)    &      2.51(9)          & 12.0 & $\nu_1+\nu_2$  \\
   4  &  1.579(1)    &      2.24(9)          &  6.1 &  \\
   5  & 51.395(2)    &      2.06(9)          & 10.3 &   \\
   6  & 53.860(2)    &      1.81(9)          & 10.3 &   \\
   7  & 47.883(2)    &      1.77(9)          &  9.6 &   \\
      &  1.450(2)    &      1.83(9)          &  5.2 & $-\nu_3+\nu_7$  \\
   8  & 32.546(2)    &      1.47(9)          &  9.0 &   \\
      & 44.130(2)    &      1.34(9)          &  9.9 & $\nu_2+\nu_5-\nu_8$  \\
   9  & 33.183(2)    &      1.26(9)          &  8.3 &   \\
      & 71.971(3)    &      1.20(9)          &  9.4 & $\nu_1+\nu_3$  \\
  10  & 45.199(3)    &      1.15(9)          &  8.7 &   \\
  11  & 33.671(3)    &      1.12(9)          &  8.4 &   \\
      & 50.691(3)    &      0.99(9)          &  6.6 & $-2\nu_4+\nu_6$  \\
      & 76.411(3)    &      0.97(9)          &  7.5 & $2\nu_1+\nu_2$  \\
      & 31.260(3)    &      0.95(9)          &  5.4 & $3\nu_8-2\nu_9$  \\
      & 73.462(3)    &      0.94(9)          &  8.1 & $\nu_1+\nu_7$  \\
      & 31.684(3)    &      0.90(9)          &  5.5 & $\nu_3-\nu_7+\nu_9$  \\
      & 41.659(3)    &      0.88(9)          &  8.4 & $-2\nu_1+2\nu_3$  \\
      & 50.262(4)    &      0.87(9)          &  5.8 & $\nu_5+\nu_8-\nu_{11}$  \\
      & 47.548(4)    &      0.86(9)          &  6.5 & $2\nu_3-\nu_{10}$  \\
      & 79.435(4)    &      0.80(9)          &  7.2 & $\nu_1+\nu_6$  \\
      & 52.553(4)    &      0.76(9)          &  5.5 & $\nu_3+\nu_5-\nu_{10}$  \\
      & 76.966(4)    &      0.76(9)          &  7.1 & $\nu_1+\nu_5$  \\
      & 40.629(5)    &      0.56(9)          &  6.1 & $-\nu_1+\nu_8+\nu_{11}$  \\
      &102.311(6)    &      0.49(9)          &  5.3 & $4\nu_1$  \\
\hline
\end{tabular}
\end{table}

\begin{table}
\small
\caption{Same as in Table\,\ref{tab:freq_TESS_s19}, but for TESS Sector 59.}
\label{tab:freq_TESS_s59}
\centering
\begin{tabular}{ccccc}
\hline
ID & $\nu$ (d$^{-1}$) & $A$ (ppt)  & S/N & Remarks\\
\hline
   1 &  25.57572(4)&     90.1(1)    &   52.4 & \\
     &  51.1513(2) &     22.3(1)    &   41.3 &  $2\nu_1$ \\
   2 &  25.2517(7) &      5.0(1)    &   19.3 & \\
     &  76.7276(9) &      3.9(1)    &   19.8 &  $3\nu_1$ \\
   3 &  47.879(1)  &      2.5(1)    &   10.3 &     \\
     &  50.827(2)  &      2.3(1)    &   10.2 & $\nu_1+\nu_2$\\
   4 &   1.460(2)  &      2.2(1)    &    7.5 & \\
     &  46.387(2)  &      2.1(1)    &   10.6 & $\nu_3-\nu_4$ \\
   5 &  32.546(2)  &      1.9(1)    &    9.8 & \\
   6 &  53.856(2)  &      1.6(1)    &    8.6 & \\
   7 &   1.294(2)  &      1.5(1)    &    5.5 & \\
   8 &  47.035(3)  &      1.5(1)    &    8.0 & \\
   9 &  33.671(2)  &      1.5(1)    &    8.5 & \\
     &  53.438(3)  &      1.4(1)    &    8.2 & $2\nu_2+2\nu_4$ \\
     &  73.454(3)  &      1.3(1)    &    8.5 & $\nu_1+\nu_3$ \\
  10 &  44.702(3)  &      1.2(1)    &    7.4 & \\
  11 &  33.176(3)  &      1.1(1)    &    6.7 & \\
     &  51.390(4)  &      1.0(1)    &    5.9 & $3\nu_4+\nu_8$ \\
     &  31.684(4)  &      0.9(1)    &    5.3 & $-\nu_4+\nu_{11}$ \\
     &   6.968(4)  &      0.9(1)    &    5.2 & $-\nu_1+\nu_5$ \\
  12 &  32.766(4)  &      0.9(1)    &    5.4 & \\
     &  31.254(4)  &      0.9(1)    &    5.1 & $\nu_5-\nu_7$\\
     &  53.026(4)  &      0.8(1)    &    5.3 & $-\nu_3+\nu_6+\nu_8$\\
     &  79.433(5)  &      0.8(1)    &    5.9 & $\nu_1+\nu_6$\\
     &  71.963(5)  &      0.8(1)    &    5.5 & $\nu_1+\nu_3-\nu_4$\\
     &  32.100(5)  &      0.8(1)    &    5.2 & $3\nu_5-2\nu_{12}$\\
     &  57.646(5)  &      0.8(1)    &    5.6 & $\nu_2-\nu_7+\nu_9$\\
     &  41.653(5)  &      0.7(1)    &    5.2 & $\nu_4-\nu_6+2\nu_8$\\
     &  76.400(5)  &      0.7(1)    &    5.2 & $2\nu_1+\nu_2$\\
     &  72.608(5)  &      0.7(1)    &    5.3 & $\nu_1+\nu_8$\\
     &  40.626(5)  &      0.7(1)    &    5.3 & $-\nu_1+\nu_5+\nu_9$\\
\hline
\end{tabular}
\end{table}

\begin{table}
\small
\caption{Same as in Table\,\ref{tab:freq_TESS_s19}, but for TESS Sector 86.}
\label{tab:freq_TESS_s86}
\centering
\begin{tabular}{ccccc}
\hline
ID & $\nu$ (d$^{-1}$) & $A$ (ppt)  & S/N & Remarks\\
\hline
   1  & 25.57557(7) &   89.6(2)    &   29.4 & \\
      & 51.1509(3)  &   22.1(2)    &   23.0 & $2\nu_1$ \\
   2  & 25.250(1)   &   5.3(2)     &   13.3 & \\
      & 76.726(2)   &   3.7(2)     &   10.9 & $3\nu_1$ \\
      & 50.825(2)   &   2.7(2)     &    7.0 &$\nu_1+\nu_2$ \\
   3  & 47.038(3)   &   2.3(2)     &    5.9 & \\
   4  & 53.856(3)   &   2.2(2)     &    6.2 & \\
   5  & 47.879(3)   &   2.0(2)     &    5.8 & \\
   6  & 50.699(3)   &   2.3(2)     &    6.2 & \\
\hline
\end{tabular}
\end{table}

\section{Mode identification for the frequencies  $\nu_1$ and $\nu_2$}
\label{sec:5}

Identification of pulsation modes from multi-colour photometry is possible because
the amplitudes and phases depend on a mode geometry which is described by
the harmonic degree, $\ell$, and azimuthal order, $m$.
In the framework of zero rotation, only the mode degree can be determined.
The linear expression for the complex amplitude of light variations
in the passband, $\lambda$, caused by pulsation in the mode with an angular frequency, $\omega$,
and geometry $(\ell,~m)$ is given by
\begin{equation}
{\cal A}_{\lambda} = {\cal D}_{\ell}^{\lambda} ({\tilde\varepsilon}
f) +{\cal E}_{\ell}^{\lambda} {\tilde\varepsilon},
\label{eq:complex_amp_ph}
\end{equation}
where
\begin{subequations}
\begin{align}
& \tilde{\varepsilon} \equiv \varepsilon Y^m_{\ell}(i,0), \\
& \mathcal{D}_{\ell}^{\lambda} = b_{\ell}^{\lambda} \frac{1}{4}
\frac{\partial \log ( \mathcal{F}_\lambda |b_{\ell}^{\lambda}| ) }
{\partial\log T_{\mathrm{eff}}}, \\
& \mathcal{E}_{\ell}^{\lambda} = b_{\ell}^{\lambda} \left[ (2+\ell
)(1-\ell ) -\left( \frac{\omega^2 R^3}{G M}
+ 2 \right) \frac{\partial \log ( \mathcal{F}_\lambda
	|b_{\ell}^{\lambda}| ) }{\partial\log g} \right],
\intertext{and}
& b_{\ell}^{\lambda} = \int_0^1 h_\lambda(\mu) \mu P_{\ell}(\mu) \, \mathrm{d}\mu.
\end{align}
\end{subequations}
Here, $\varepsilon$ denotes the intrinsic mode amplitude, $i$ is the inclination angle, $f$ is the non-adiabatic parameter that gives the ratio
of the relative flux variation to the relative radial displacement of the surface at the level of photosphere.
Both $\varepsilon$ and $f$ have to be regarded as complex.
Then, $h_\lambda(\mu)$ is the limb darkening law, $P_\ell(\mu)$
is the Legendre polynomial and $G,~M, ~R$ have their usual meanings.
The values of $b_{\ell}^{\lambda}$ and the partial derivatives
of ${\cal F}_\lambda |b_{\ell}^{\lambda}|$ have to be calculated from model atmospheres.
The amplitude of the brightness changes is given by $A_{\lambda}=\sqrt{A_R ^2+A_I^2}$
and the phase by $\varphi_{\lambda}=\arctan(A_I/A_R)$, where $A_R$ and $A_I$ are the real and imaginary parts of $A$ in Eq.\,\ref{eq:complex_amp_ph}.

In the linear approximation of pulsations, the mode identification can be performed either using amplitude ratios and phase differences in different bands
or by simultaneously determining the mode degree, $\ell$, and the intrinsic amplitude multiplied by $Y_\ell^m(i,0)$.
Regarding the non-adiabatic parameter $f$, two approaches can be adopted: in the first approach,
its values for a given mode are taken from pulsational computations, while in
the second approach, they are determined from multi-colour amplitudes and phases, simultaneously with $\tilde \varepsilon$ \citep{2003A&A...407..999D,2005A&A...441..641D}.

In our pulsational computations, we relied on the non-adiabatic code of W.\,Dziembowski \citep{Dziembowski1977,Pamyatnykh1999}, which
includes the effects of rotation on pulsational frequencies within the perturbation approach, taking into account non-spherically symmetric
distortion due to the centrifugal force and the second- and third-order effects of the Coriolis force.
The convective flux freezing approximation is used, which is justified if convection is not very efficient in the envelope.

In the case of BL~Cam, we used two approaches based on the $BVI$ time-series photometry of
\cite{2007A&A...471..255R}.

\subsection{Photometric diagnostic diagrams for SX Phe star models}
\begin{figure}[htb!]
	\includegraphics[width=\columnwidth,clip]{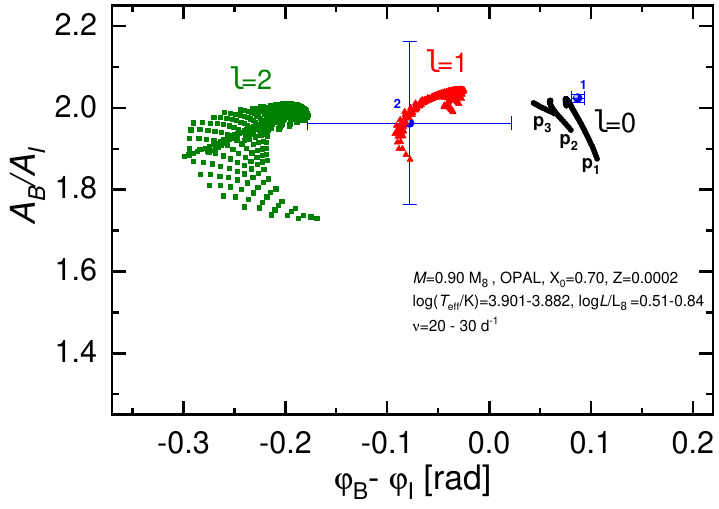}
	\includegraphics[width=\columnwidth,clip]{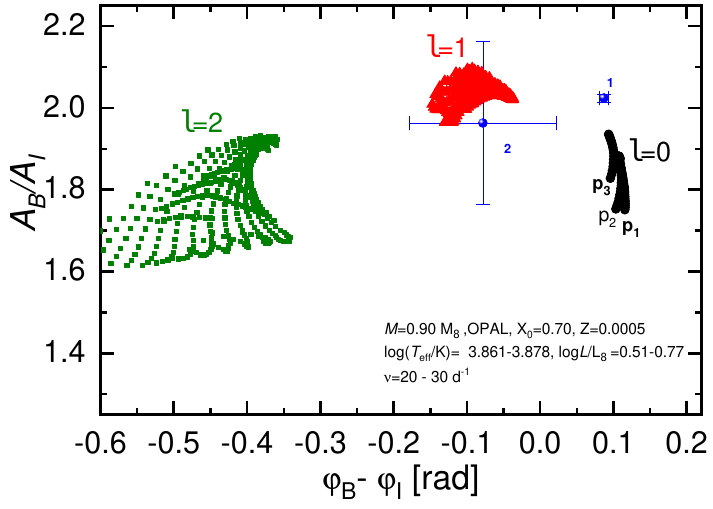}
	\includegraphics[width=\columnwidth,clip]{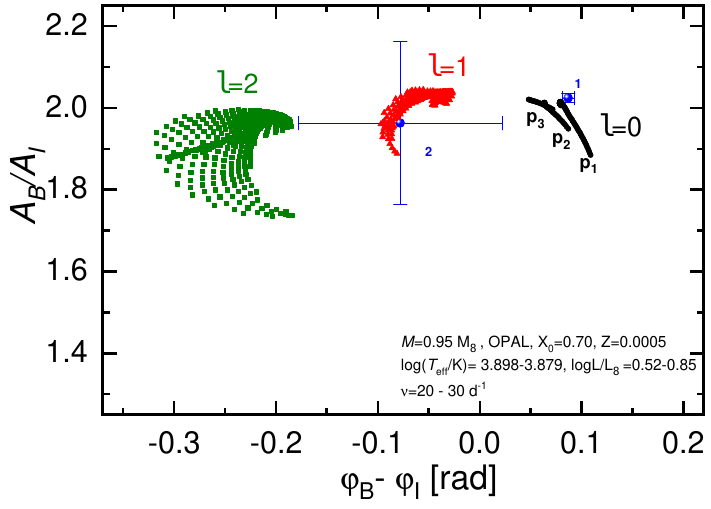}
	\caption{Positions of the frequencies $\nu_1$ and $\nu_2$ on the $BI$ photometric diagram showing the amplitude ratio vs the phase difference. The models have an initial hydrogen abundance of $X_0=0.70$
		and mixing length parameter of $\alpha_{\rm MLT}=0.5$. Models shown for a mass of $M=0.90$\,M$_{\sun}$ and metallicity of $Z=0.0002$ (top), for $M=0.90$\,M$_{\sun}$ and $Z=0.0005$ (middle), and for $M=0.95$\,M$_{\sun}$ and $Z=0.0005$ (bottom).
		Kurucz ODFNEW atmosphere models  were adopted with the metallicity [m/H]$=-2.0,~-1.5$, and micro-turbulent velocity of $\xi_t=2\,\mathrm{km\,s^{-1}}$.}
		\label{ModeID}
\end{figure}

Generating diagrams with amplitude ratios and phase differences in various pairs of photometric passbands
is one of the most popular and transparent ways to visualise the identification of pulsation modes.
Up to now, there have been no published results of this type for SX Phe stars in the literature.
In general, the greater the difference in wavelength between photometric passbands,
the better the separation of modes with various degrees $\ell$ in the photometric diagnostic diagrams.
In Fig.\,\ref{ModeID}, we show examples of photometric diagrams in the $BI$ filters for models of BL~Cam.
The effects of metallicity (top vs middle panels) and  mass
(middle vs bottom panels) are illustrated. Models with $(T_{\rm eff},~L)$ values within the error box were considered
and only modes with frequencies in the range (20,~30)\,d$^{-1}$ were included.
The flux derivatives with respect to $\log T_{\rm eff}$ and $\log g$, limb darkening $h_\lambda(\mu)$ and its derivatives were determined from Kurucz ODFNEW atmosphere models
with the metallicity [m/H]$=-2.0$ for $Z=0.0002$ and [m/H]$=-1.5$ for $Z=0.0005$, and with the micro-turbulent velocity of $\xi_t=2\,\mathrm{km\,s^{-1}}$.

The observed values for the frequencies $\nu_1$ and $\nu_2$ are plotted along with the errors.
As we can see, for the dominant frequency, $\nu_1$, the identification is uniquely $\ell=0$ with an indication of the fundamental mode $(p_1)$.
In the case of the second frequency, $\nu_2$, the errors are much larger. Nevertheless, we can conclude with fairly high confidence that $\nu_2$
is likely a dipole mode.

\subsection{Simultaneous determination of $\ell$, $f$ and $\varepsilon$ for the dominant frequency}

In this approach, the system of $N$ complex equations (1) is solved for a given $\ell$ and $(T_{\rm eff},\log g)$ to determine $\tilde\varepsilon$ and $f$. Here, $N$ is the number of passbands. The degree $\ell$ and associated complex values of $\tilde\varepsilon$ and $f$ are considered as most probable if there is a clear minimum in the difference between the calculated and observed photometric amplitudes and phases.
The discriminant, which gives a goodness of fit, is defined as
\begin{equation}
\chi^2=\frac1{2N-N_p} \sum_{i=1}^N  \frac{ \left|{\cal A}^{obs}_{\lambda_i} - {\cal A}^{cal}_{\lambda_i}\right|^2 }
{ |\sigma_{\lambda_i}|^2},
\end{equation}
where the superscripts $obs$ and $cal$ denote the observed and calculated complex amplitudes, respectively.
$N_p$ is the number of parameters to be determined and $N_p=4$ because there are two complex parameters, $\tilde\varepsilon$ and $f$. The observational errors, $\sigma_{\lambda}$, are computed as
\begin{equation}
|\sigma_\lambda|^2= \sigma^2 (A_{\lambda})  +  A_{\lambda}^2 \sigma^2(\varphi_\lambda),
\end{equation}
where $\sigma  (A_{\lambda})$ and $\sigma (\varphi)$ are the errors of observed amplitudes and phases in a passband $\lambda$, respectively.

\begin{figure}
	\includegraphics[width=88mm,clip]{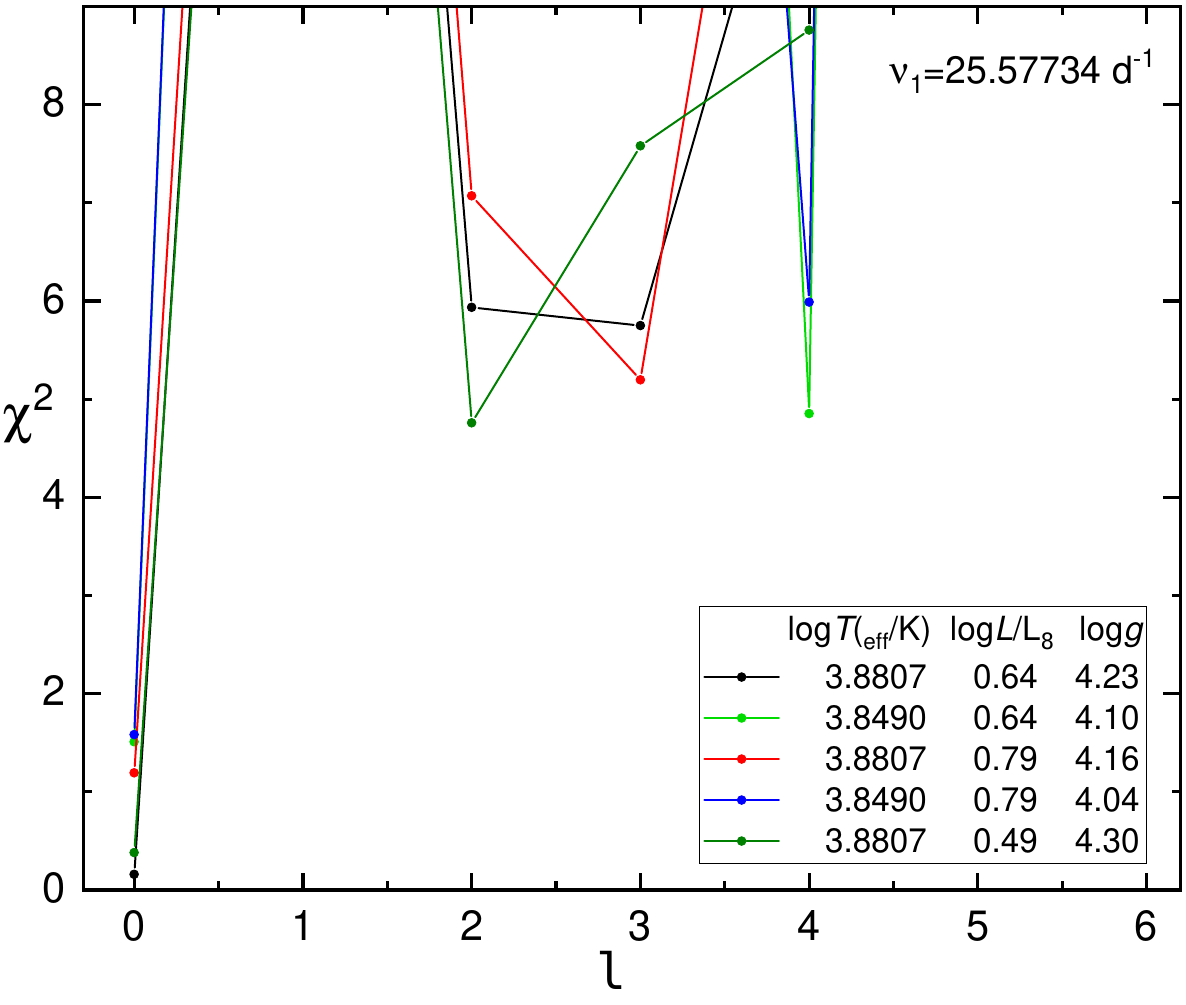}
	\caption{Discriminant $\chi^2$ as a function of the mode degree $\ell$ for the dominant frequency of BL~Cam. Kurucz ODFNEW atmosphere models were adopted with the metallicity [m/H]$=-2.0$ and micro-turbulent velocity $\xi_t=2\,\mathrm{km\,s^{-1}}$.}
\label{ModeID2}
\end{figure}

The values of $\chi^2$ as a function of degrees $\ell$ for the dominant frequency of BL~Cam are shown in Fig.\,\ref{ModeID2}.
We assumed the parameters $(T_{\rm eff},~L)$ from the whole error box. As one can see, the second approach also leads
to the conclusion that $\nu_1$ is undoubtedly a radial mode.
In the case of $\nu_2$, however, the large uncertainties in amplitudes and phases prevent a reliable mode identification using the second approach.

\begin{figure}
	\includegraphics[width=88mm,clip]{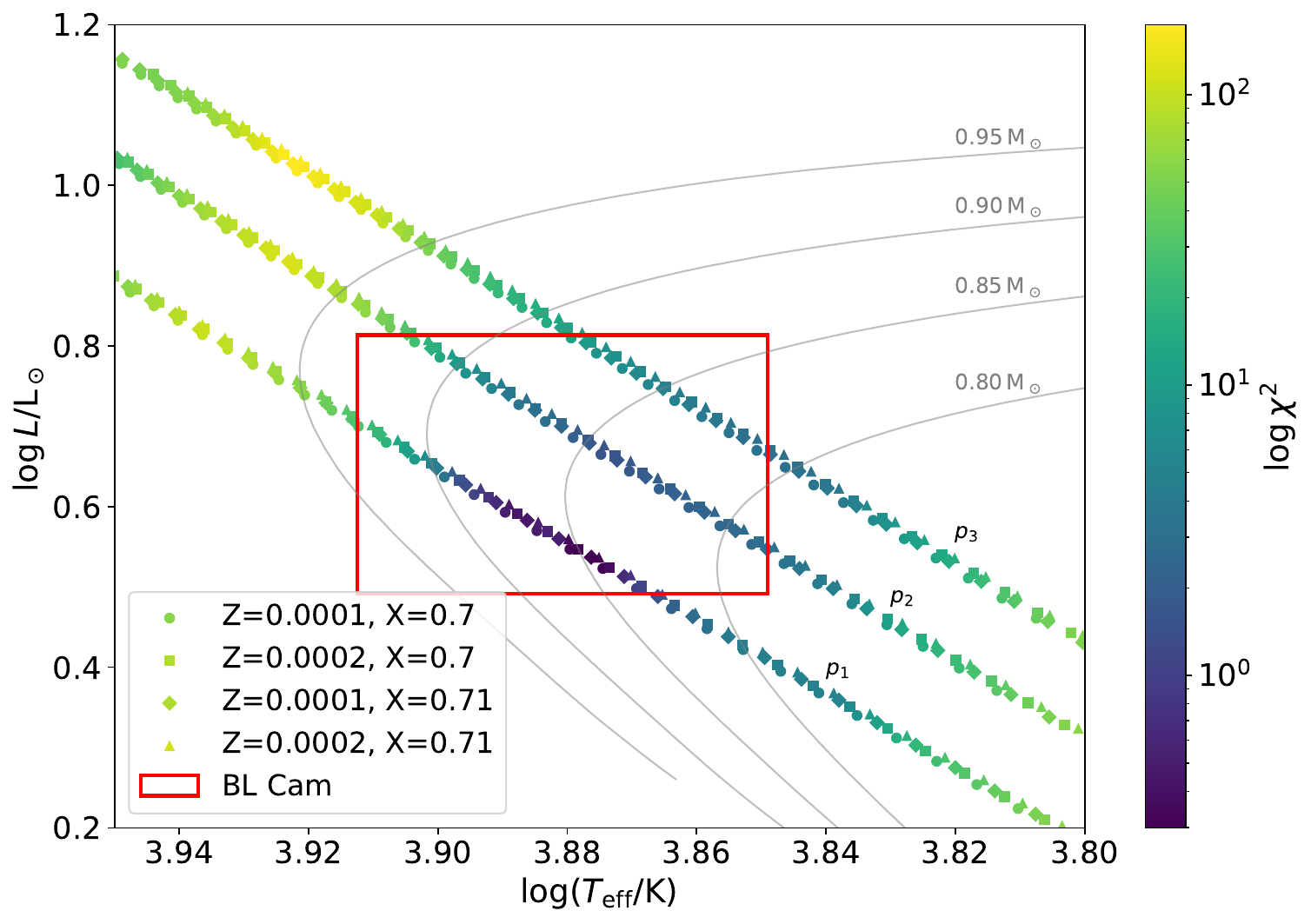}
	\caption{HRD with the position of BL~Cam indicated as well as the lines of a constant frequency $\nu=25.57734$ of the dominant radial mode, considered as fundamental ($p_1$), first-overtone ($p_2$), and second-overtone ($p_3$).
		The symbols represent different values of the initial hydrogen abundance, $X_0$, and metallicity, $Z$.
		The values of $\chi^2$, defined in Eq.\,(3), are coded with greenish colours. }
\label{ModeID3}
\end{figure}

\begin{figure}
	\includegraphics[width=88mm,clip]{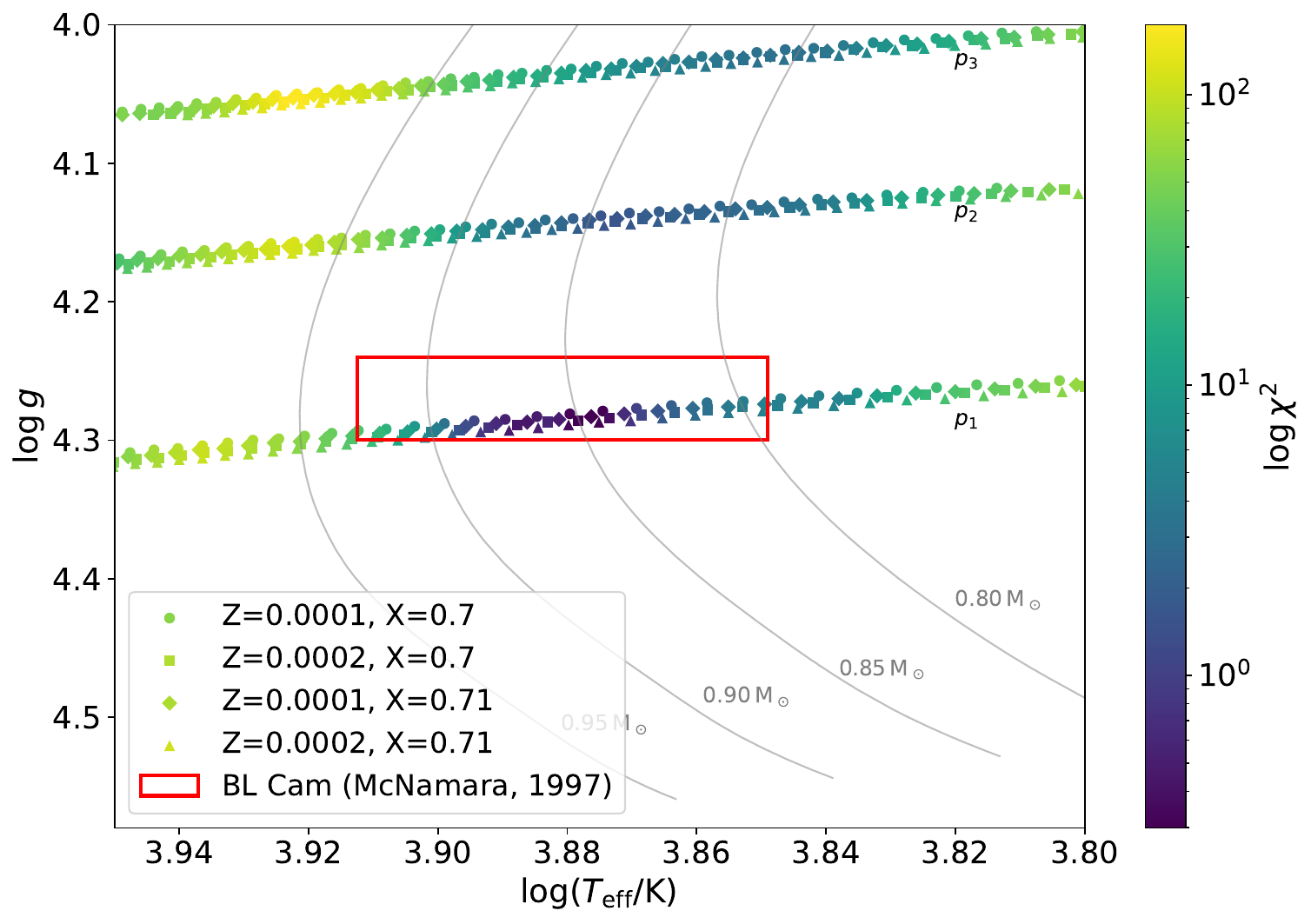}
	\caption{Same result as in Fig.\,\ref{ModeID3} but translated to the Kiel diagram.}
\label{ModeID4}
\end{figure}

In the case of radial modes, we can choose certain parameters to move along the lines of constant frequencies on the HRD. Therefore, in the next step, we calculated the values of $\chi^2$ assuming the hypothesis of a fundamental radial mode, $p_1$,
the first overtone, $p_2$, and the second overtone, $p_3$. These lines are depicted along with the error box in the HRD in Fig.\,\ref{ModeID3}.
The values of $\chi^2$ are colour coded. We can see that all three lines are inside the error box, but the lowest values of $\chi^2$ are achieved for the fundamental mode. Since BL~Cam has a well-determined surface gravity by \cite{1997PASP..109.1221M},
we translated the HRD into the Kiel diagram shown in Fig.\,\ref{ModeID4}. The values of $\log g$ unambiguously indicate that $\nu_1$
is the radial fundamental mode.
The same result was obtained by \cite{2007A&A...471..255R}.

To further support this conclusion, we estimated the pulsation constant, $Q = P\sqrt{\rho/\rho_{\odot}}$,
for the dominant frequency $\nu_1=25.57734$\,d$^{-1}$. Instead of using the transformation
given by \citet{1990A&A...231...56B} and widely used in the literature, we applied an equivalent relation
expressed in therms of the luminosity, $L$, rather than the bolometric magnitude, $M_{\mathrm{bol}}$.
Applying the elementary transformation,
$$Q = P \left(\frac{g}{g_{\odot}}\right)^{1/2} \left(\frac{L}{L_{\odot}}\right)^{-1/4}
\left(\frac{T_{\rm eff}}{T_{\rm eff,\odot}}\right),$$
and adopting the solar values  $T_{\rm eff,\odot}=5772$\,K and $\log g_{\odot}=4.438$\,dex,  we obtained
$$\log Q=\log P+\frac12 \log g +\log T_{\rm eff}-\frac14 \log L/L_{\odot}-5.9803.$$
Using the central values of $\log T_{\rm eff}=3.881$,  $\log L/L_{\odot}=0.65$ (derived in Sect. 2),
together with $\log g=4.3$ \citep{1997PASP..109.1221M}, we found a pulsation constant of $Q=0.0302$\,d.
This value of $Q$ is consistent with pulsation in the fundamental radial mode for SX Phe-type stars.
It supports the interpretation based on the Kiel diagram shown in Fig.\,\ref{ModeID4} and is also consistent
with the results from the $BI$ photometric diagram (Fig.\,\ref{ModeID}), where the top
and bottom panels both indicate a fundamental mode.

We also derived the pulsation constant, $Q$, for the frequency of $\nu_8=32.546$\,d$^{-1}$,
obtaining $Q=0.0238$\,d. This value is in agreement with the theoretical range for the radial first overtone.
Crucially, the corresponding period ratio of $P_8/P_1=0.7859$ closely matches the theoretical ratio
expected for radial fundamental and first-overtone pulsators, further strengthening our mode identification.


\section{Seismic models of BL~Cam}
\label{sec:6}

We performed seismic modelling of BL~Cam using a Monte Carlo-based Bayesian analysis.
The procedure follows closely the approach presented by \citet{2022MNRAS.512.3551D, 2023MNRAS.526.1951D, 2025MNRAS.539.3381D}.
For each stellar model, we calculated the corresponding pulsation spectrum and compared the theoretical predictions with the available observational constraints. The likelihood was evaluated adopting the standard form,
\begin{equation}
\mathcal{L}(E|H) = \prod_{i=1}^{N} \frac{1}{\sqrt{2\pi\sigma_i^2}}
\exp\left[-\frac{\left(\mathcal{O}_i-\mathcal{M}_i\right)^2}{2\sigma_i^2}\right],
\end{equation}
where $\mathcal{O}_i$ and $\mathcal{M}_i$ denote the observed and modelled values of the $i$-th observable, respectively, and $\sigma_i$
is the corresponding observational uncertainty.
The hypothesis vector $H$ includes the adjustable stellar parameters, namely, the mass, $M$,
the initial hydrogen abundance, $X_0$, metallicity, $Z$, the mixing-length parameter, $\alpha_\mathrm{MLT}$,
and the initial rotational velocity, $V_{\mathrm{rot},0}$.
The evidence vector, $E$, consists of the selected seismic and non-seismic constraints,
including the frequencies of identified pulsation modes, and the non-adiabatic $f$ parameter, as well as the luminosity and effective temperature.

All evolutionary models were computed with the Warsaw-New Jersey evolutionary code, adopting OPAL opacity tables. We assumed a single-star evolutionary scenario, since there is currently no observational evidence indicating that BL~Cam is the product of mass transfer, a merger event, or a stellar collision; furthermore, the long orbital period, $P_\mathrm{orb}\approx 144$\,d, also argues against recent binary interaction.
Moreover, once such an object is formed and reaches thermal equilibrium, its subsequent evolution is expected
to resemble that of a single star of comparable mass. The non-adiabatic parameter, $f$, was determined using the system
of Equations\,\ref{eq:complex_amp_ph}
for the $BVI$ amplitudes and phases derived from the photometry of \citet{2007A&A...471..255R}.

For each mode-identification hypothesis considered in the following subsections,
the Bayesian analysis was based on approximately $5\times10^{5}$ iterations,
each requiring the computation of a new stellar model and its corresponding pulsation frequencies and amplitudes.

\subsection{Fitting two modes: Radial and dipole}
\label{Sec:2modes}
We began the seismic modelling by fitting the two well-identified pulsation modes (discussed in Section\,\ref{sec:5}),  $\nu_1$ and $\nu_2$.
The dominant frequency, $\nu_1$, was identified as the radial fundamental, while $\nu_2$ was interpreted as a dipole mode. In addition to the pulsation frequency, we also fit the non-adiabatic $f$ parameter of the radial mode. Since the azimuthal order $m$ of the dipole mode is unknown,
we considered all three possible values (i.e. $m=-1,0,$ and $1$). The radial order of the dipole mode was treated as a free parameter and fitted simultaneously.

\begin{table*}
\small
\caption{Median parameter values of the OPAL seismic models of BL~Cam derived from Monte Carlo simulations.}
\label{tab:Bayes_results}
\centering
\renewcommand{\arraystretch}{1.4}
\begin{tabular}{cccccccccc}
\hline
 $M$                   &$X_0$&$Z$&$\log\left(T_\mathrm{eff}/\mathrm{K}\right)$&$\log L/\mathrm L_{\sun}$&$\log g$&$V_{\mathrm{rot},0}$     &$V_{\mathrm{rot}}$       &$\alpha_\mathrm{MLT}$&     age  \\
  $(\mathrm M_{\sun})$ &     &   &                                            &                         & (dex)  &$(\mathrm{{km\,s}^{-1}})$&$(\mathrm{{km\,s}^{-1}})$&                     &     (Gyr)    \\
 \hline
\multicolumn{10}{c}{\underline{\textit{Fitted modes:} $\ell=0\,p_1$ and $\ell=1,\,m=0$ }} \\
 $0.961^{+0.002}_{-0.002}$ &  $0.7012^{+0.0005}_{-0.0009}$ &  $0.0004775$ &  $3.8979^{+0.0005}_{-0.0005}$ &  $0.655^{+0.002}_{-0.003}$ &  $4.2920^{+0.0004}_{-0.0007}$ &  $76.2^{+0.8}_{-0.3}$ &  $94.6^{+1.0}_{-0.4}$ &  $0.2^{+0.1}_{-0.1}$ & $3.84^{+0.02}_{-0.02}$   \\
 \hline
 \multicolumn{10}{c}{\underline{\textit{Fitted modes:} $\ell=0\,p_1$ and $\ell=1,\,m=+1$ }} \\
 $0.954^{+0.002}_{-0.003}$ &  $0.6987^{+0.0002}_{-0.0003}$ &  $0.0002932$ &  $3.9148^{+0.0007}_{-0.0007}$ &  $0.724^{+0.003}_{-0.004}$ &  $4.3019^{+0.0006}_{-0.0006}$ &  $25.1^{+4.3}_{-4.5}$ &  $32.4^{+5.6}_{-5.9}$ &  $1.5^{+0.7}_{-0.9}$ & $4.05^{+0.03}_{-0.03}$  \\
 \hline
 \multicolumn{10}{c}{\underline{\textit{Fitted modes:} $\ell=0\,p_1$ and $\ell=1,\,m=-1$ }} \\
 $0.950^{+0.001}_{-0.002}$ &  $0.7004^{+0.0002}_{-0.0003}$ &  $0.0003890$ &  $3.8985^{+0.0003}_{-0.0004}$ &  $0.654^{+0.001}_{-0.002}$ &  $4.2858^{+0.0002}_{-0.0003}$ &  $84.3^{+0.1}_{-0.1}$ &  $106.0^{+0.1}_{-0.2}$ &  $0.2^{+0.2}_{-0.1}$ &  $4.01^{+0.02}_{-0.01}$  \\
  \hline
  \multicolumn{10}{c}{\underline{}} \\
 \multicolumn{10}{c}{\underline{\textit{Fitted modes:} $\ell=0\,p_1$; $\ell=0\,p_2$ and $\ell=1,\,m=0$ }} \\
 $0.959^{+0.004}_{-0.002}$ &  $0.7011^{+0.0007}_{-0.0005}$ &  $0.0004735$ &  $3.8973^{+0.0010}_{-0.0007}$ &  $0.652^{+0.005}_{-0.003}$ &  $4.2913^{+0.0007}_{-0.0005}$ &  $76.7^{+0.3}_{-0.4}$ &  $95.2^{+0.4}_{-0.4}$ &  $0.2^{+0.1}_{-0.1}$ &  $3.86^{+0.03}_{-0.04}$ \\
  \hline
 \multicolumn{10}{c}{\underline{\textit{Fitted modes:} $\ell=0\,p_1$; $\ell=0\,p_2$ and $\ell=1,\,m=-1$ }} \\
$0.978^{+0.002}_{-0.003}$ &  $0.7056^{+0.0002}_{-0.0005}$ &  $0.0004157$ &  $3.9051^{+0.0005}_{-0.0008}$ &  $0.689^{+0.002}_{-0.004}$ &  $4.2939^{+0.0003}_{-0.0006}$ &  $77.53^{+0.08}_{-0.12}$ &  $96.5^{+0.2}_{-0.1}$ &  $0.6^{+0.3}_{-0.4}$ &  $3.74^{+0.03}_{-0.02}$  \\
\hline
\end{tabular}
\tablefoot{In all cases, the inferred error in $Z$ was $\leq 3 \times 10^{-7}$.}
\end{table*}

\begin{figure}
	\includegraphics[width=\columnwidth,clip]{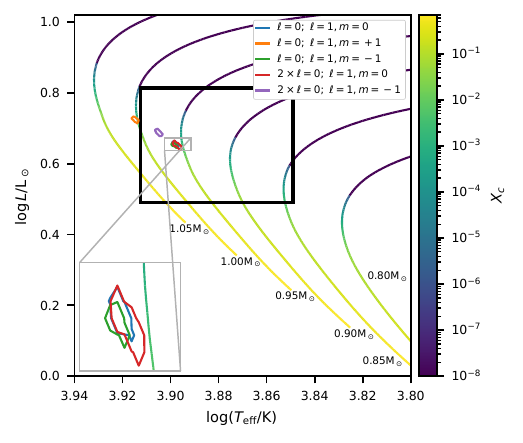}
	\caption{Position of seismic models of BL~Cam in the HRD. For reference, evolutionary tracks are computed with the OPAL opacity tables; the AGSS09 solar mixture, the mixing-length parameter, $\alpha_{\rm MLT}=0.2$; an initial rotation velocity, $V_{{\rm rot},0}=77\,\mathrm{km\,s^{-1}}$; metallicity, $Z = 0.00047$; and  initial hydrogen abundance, $X_0 = 0.7$. The parameters correspond to values listed in Table\,\ref{tab:Bayes_results}. The colours of evolutionary tracks represent the values of central hydrogen abundance, $X_c$.}
\label{fig:HR_seismic_models}
\end{figure}

The results of our Bayesian seismic modelling are summarised in Table\,\ref{tab:Bayes_results} (first three rows).
Independently of the assumed azimuthal order $m$ of the dipole mode, the inferred stellar mass is consistently close to $0.95\,\mathrm{M}_{\sun}$.
A similar conclusion applies to the initial hydrogen abundance, which is found to be $X_0 \approx 0.7$ in all cases.
The inferred stellar age is also comparable for all solutions, with $t \approx 4$\,Gyr.
We also note that in all seismic models, the fitted pulsation modes are predicted to be unstable.
The stability of a mode is quantified by the parameter $\eta$, defined as the normalised work integral \citep{1978AJ.....83.1184S}.
Positive values of $\eta$ correspond to excited modes, whereas negative values indicate damped modes. For all fitted modes considered here, we find $\eta>0$.
Furthermore, we find that the dipole mode is $g_4$, $g_5$, and $g_3$ for the axisymmetric, prograde, and retrograde cases, respectively.

In contrast, significant differences are observed in the inferred metallicity.
For simulations assuming an axisymmetric dipole mode, we obtain $Z \approx 0.00048$.
In the case of prograde modes, a substantially lower metallicity of $Z \approx 0.00029$ is inferred,
while for retrograde modes we derive an intermediate value of $Z \approx 0.00039$.
Clear differences are also seen in the rotational velocities.
For models assuming the dipole mode to be axisymmetric or retrograde, the current surface rotation velocity is found to be
$\sim 95$ and $106\,\mathrm{km\,s^{-1}}$, respectively, whereas
the prograde solutions yield a much lower rotation velocity of approximately $32\,\mathrm{km\,s^{-1}}$.
Differences are likewise evident in the mixing-length parameter $\alpha_{\rm MLT}$.
For the axisymmetric and retrograde solutions, $\alpha_{\rm MLT}$ is close to $0.2$,
whereas in the prograde case it reaches significantly higher values, around $1.5$,
with a considerably broader posterior distribution (cf. Fig.\ref{fig:hist_B1}, top-middle panel).
Posterior probability distributions of all fitted parameters are presented in Appendix\,\ref{app2}
(Figs\,\ref{fig:hist_B1} and \ref{fig:hist_B2}).

The positions of the seismic models in the HRD are shown in Fig.\,\ref{fig:HR_seismic_models}.
For the hotter models (solution for a dipole prograde mode), the theoretical amplitudes reproduce the observed ones less accurately.
This discrepancy is visible in Figs.\,\ref{ModeID3} and\,\ref{ModeID4},
and is additionally reflected in the larger uncertainties of the derived $f$ parameter.

In the case of radial modes, $\tilde\varepsilon$
derived from the set of Eq.\,\ref{eq:complex_amp_ph} is equal to the intrinsic mode amplitude, $\varepsilon$ (i.e.
the relative radial displacement amplitude at the stellar surface). Therefore, we were also able to determine this paramerer for $\nu_1$, obtaining
$0.01015^{+0.00004}_{-0.00004}$ for the axisymmetric dipole mode scenario;
$0.00967^{+0.00002}_{-0.00001}$ for the prograde case; and
$0.01006^{+0.00003}_{-0.00002}$ for the retrograde case.

\subsection{Fitting three modes: Two radial and dipole}

\begin{figure*}
	\includegraphics[width=\textwidth,clip]{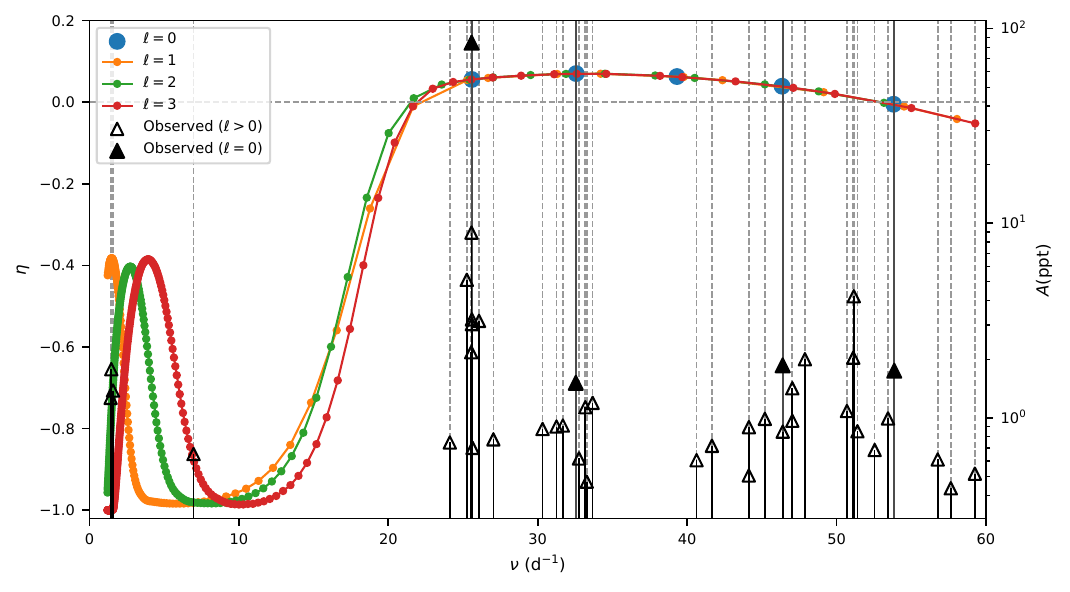}
	\caption{Instability parameter, $\eta$, as a function of frequency for the seismic model computed at $\log T_{\rm eff}=3.89718$. The remaining stellar parameters are adopted from the solution based on two radial modes and one dipole axisymmetric mode, with the median values listed in Table\,\ref{tab:Bayes_results}. The values of the instability parameter, $\eta$, are read from the left-hand $y$-axis, while the right-hand $y$-axis shows the observed amplitudes, $A$. The observed frequencies corresponding to radial (and probable radial) modes are marked by solid vertical lines and filled triangles, whereas non-radial modes are indicated by dashed vertical lines and open triangles. The height of each triangle corresponds to the observed pulsation amplitude. The vertical lines are drawn across the entire plot to facilitate comparison between the observed and theoretical frequencies. Theoretical mode frequencies are shown as coloured symbols located at the corresponding values of $\eta$ and connected by lines. Different colours denote different spherical degrees: $\ell=0$ (blue), $\ell=1$ (orange), $\ell=2$ (green), and $\ell=3$ (red).}
		\label{fig:best_model}
\end{figure*}

We repeated the simulations described in Section\,\ref{Sec:2modes}, but including an additional mode, namely, the probable second radial mode, expressed as
$\nu_8 = 32.546$\,d$^{-1}$, which would correspond to the first overtone.
This time we were able to reproduce all fitted parameters only under the assumption that the dipole mode is either axisymmetric or retrograde.
As in the case of the two-frequency fit, we find that the dipole mode corresponds to $g_4$  and $g_3$ for the axisymmetric or retrograde cases, respectively.
The results of our Bayesian analysis are
summarised in Table\,\ref{tab:Bayes_results} (the last two rows), while the positions of the seismic models are shown in the HRD in Fig.\,\ref{fig:HR_seismic_models}.

For the axisymmetric case, the inferred parameters are similar to those obtained in the two-mode simulations. In contrast, for the retrograde assumption, we observed
noticeable differences. In particular, models resulting from fitting three modes are hotter compared to those obtained from the two-mode fit.
We also derived a slightly higher value of $\alpha_\mathrm{MLT}$ (with a broader posterior distribution) and a higher metallicity. The higher $T_\mathrm{eff}$
leads to a poorer  reproduction of the observed amplitudes and phases of the radial fundamental mode and is reflected in larger uncertainties of the $f$ parameter.
This suggests that the retrograde solution is disfavoured.
We also note that for both scenarios almost all fitted modes (all but a negligible fraction among several hundred thousand models)
are excited (i.e. $\eta>0$).

In Fig\,\ref{fig:best_model}, we depict the instability parameter $\eta$ (the left-hand $y$-axis)
as a function of the mode frequency for a representative seismic model in which
we fit the dipole mode as axisymmetric.
The right-hand y-axis of Fig.\,\ref{fig:best_model} corresponds to the TESS amplitudes of
the observed frequencies. For clarity, we omitted the highest independent frequency, $\nu_{68}=76.26867$\,d$^{-1}$ (see Table\,\ref{tab:freq_TESS_combined}).
This mode is stable in our model and cannot be explained within the present framework. Moreover, it has one of the smallest amplitudes and its S/N ratio is close to our detection threshold.

Importantly, the main variability region, where the largest number of frequencies is detected and the amplitudes are highest, is well covered by unstable radial and non-radial modes. The instability parameter, $\eta$, is positive for frequencies between approximately 21 and 53\,d$^{-1}$.
Only a few modes near 60\,d$^{-1}$ have a slightly negative $\eta$ (see Fig.\,\ref{fig:best_model}).

In addition, the model predicts the third and fourth radial overtones at frequencies close to $\nu_9=46.39032\,\mathrm{d}^{-1}$
and $\nu_{12}=53.85822\,\mathrm{d}^{-1}$ (Table\,\ref{tab:freq_TESS_combined}),
previously reported as f$_{19}$ and f$_3$, respectively, with ground-based observations in \cite{2007A&A...471..255R}.
These observed frequencies may therefore correspond to the radial modes. These observed frequencies are marked by filled triangles in
Fig\,\ref{fig:best_model}. The mode corresponding to frequency $\nu_{12}$ is only marginally stable ($\eta=-0.005$); therefore, even a small change in the model parameters could shift it to the unstable domain. All other detected radial modes are unstable.

Interestingly, the detected frequency set lacks a second radial overtone.
However, in the periodogram, we note a weak signal with $\mathrm{S/N}=3.9$ at 39.304\,d$^{-1}$, which does not satisfy our adopted detection criterion, but is close to the theoretical value predicted for the second overtone in our model.

In the vicinity of the fundamental radial mode, we observe a dense oscillation spectrum. Our model also predicts closely spaced $\ell=1,\,2$ and 3 modes in this region.
Such a dense spectrum may facilitate amplitude modulation effects, including beating between closely spaced modes, as well as possible nonlinear mode interactions.
In Fig.\,\ref {fig:best_model} we show only axisymmetric modes, since including rotational splitting would make the oscillation spectrum much denser. Moreover, we cannot exclude
the presence of higher degree-modes, as they are also predicted to be unstable.

In the low-frequency domain, our model does not predict instability. Nevertheless, we observed a local maximum of $\eta$ in this region, particularly around frequencies
near 1.5\,d$^{-1}$. It should be noted that in our pulsation calculations only the $\kappa$ excitation mechanism is included.
However, by analogy with $\gamma$~Dor stars, an additional driving mechanism may operate, such as the convective flux blocking near the base of the convective envelope
or coupling between convection and pulsations \citep[e.g.][]{2000ApJ...542L..57G, 2005A&A...435..927D, 2016MNRAS.457.3163X}.
We note that the theoretical oscillation spectrum in this region is also very dense.

Similarly to the two-frequency fit, we also derived the intrinsic mode amplitude. We obtained
$0.01019^{+0.00005}_{-0.00008}$ assuming an axisymmetric dipole mode, and
$0.00966^{+0.00002}_{-0.00002}$ for the retrograde case.
Regardless of the adopted scenario (i.e. two- or three-frequency fit, as well as the assumed azimuthal order of the dipole mode), we find that for BL~Cam,
the relative radial displacement amplitude at the stellar surface for the $\nu_1$ mode is of the order of  1 per cent.

The posterior distributions for all five  mode identification cases are presented in Appendix\,\ref{app2} (Figs.\,\ref{fig:hist_B1} and \ref{fig:hist_B2}).

\section{Summary}
\label{sec:7}

BL~Cam is an extremely low-metallicity ($Z\lesssim 0.0005$) SX Phoenicis star, making it a particularly valuable laboratory for testing pulsation and evolutionary models in the low-$Z$ regime.
In this work, we analysed new photometric data from ZTF and TESS in combination with detailed seismic modelling.

The long time-base ZTF data confirm the binary nature of BL~Cam and allow us to redetermine the orbital period to approximately 144\,d.
Combining the seismic mass, $M_\mathrm{A}\approx 0.96\,\mathrm{M}_\sun$,
derived from our modelling with the mass function derived from ZTF data, we were able to estimate the minimum mass of the companion as $M_\mathrm{B,\,min} \approx 0.46\,\mathrm{M}_{\odot}$.
The high-precision TESS photometry enabled the detection of numerous pulsation modes. Not all frequencies previously reported by \citet{2007A&A...471..255R}
were recovered, which could indicate large-amplitude variations over time.

Using Bayesian seismic modelling, we derived very tight constraints on the stellar parameters. We analysed two configurations: a fit to the frequencies of two dominant modes
($\ell=0$ and $\ell=1$) and an extended solution including
the third frequency as the first radial overtone.

For the two-mode case, the inferred stellar mass, age, and initial hydrogen abundance are remarkably robust with respect to the assumed azimuthal order of the dipole mode.
In all scenarios, the stellar mass is close to $0.95\,\mathrm{M}_{\sun}$, the initial hydrogen abundance is $X_0\approx 0.7$, and the age is about\,4 Gyr.
However, significant differences were found in the metallicity, rotation rate, and mixing-length parameter, which remain sensitive to the assumed mode geometry.
All fitted modes were predicted to be unstable ($\eta>0$).

When the additional radial overtone is included, acceptable solutions are obtained only for axisymmetric or retrograde dipole identification. The retrograde case leads to
systematically hotter models and a broader distribution of $\alpha_\mathrm{MLT}$. These models also reproduce the observed amplitude and phase
of the fundamental radial mode less accurately, resulting in larger uncertainties in the $f$ parameter. This might favour a model in which the dipole mode is axisymmetric.

Our seismic models reproduce the main variability domain well, where most of the detected frequencies
and the highest amplitudes are observed. In this frequency range, the instability parameter $\eta$ is positive.
The representative model predicts higher radial overtones with frequencies close to $\nu_9 = 46.39032\,\mathrm{d}^{-1}$ and $\nu_{12} = 53.85822\,\mathrm{d}^{-1}$,
suggesting that two additional radial modes may be present in the oscillation spectrum of BL~Cam.
More data are needed to confirm this mode identification and to enable further extended seismic modelling.

We detected several low-frequency signals, which suggest that BL~Cam may be a hybrid pulsator exhibiting both low-order $p$ and $g$ modes as well as high-order $g$ modes.
However, in the low-frequency domain, the present models do not predict instability within the framework of the $\kappa$ mechanism alone. Additional driving processes,
such as convective flux blocking or convection-pulsation coupling, similar to the effects at work in $\gamma$~Dor stars, might therefore be required. A detailed treatment of these effects is beyond the scope of this work.

\begin{acknowledgements}
The work was financially supported by the Polish National Science Centre grant 2023/50/A/ST9/00144.

ER acknowledges financial support from the Spanish Agencia Estatal de
Investigación (AEI/10.13039/501100011033) of the Ministerio de Ciencia e
Innovación through projects PID2022-137241NB-C43 and
PID2023-149439NB-C42, and the Centre of Excellence 'Severo Ochoa'
award to the Instituto de Astrofísica de Andalucía (grant CEX2021-001131-S
funded by MCIN/AEI/10.13039/501100011033).

Calculations were carried out using resources provided by Wrocław Centre for Networking and
Supercomputing (\url{http://wcss.pl}), grant no. 265.

This work has made use of data from the European Space Agency
(ESA) mission Gaia (\url{https://www.cosmos.esa.int/gaia}), processed
by the Gaia Data Processing and Analysis Consortium (DPAC; \url{https://www.cosmos.esa.int/web/gaia/dpac/consortium}). Funding for the
DPAC has been provided by national institutions, in particular the
institutions participating in the Gaia Multilateral Agreement.

This paper includes data collected by the \textit{TESS} mission. Funding
for the \textit{TESS} mission is provided by the NASA Explorer Program.

Based on observations obtained with
the Samuel Oschin Telescope 48-inch and the 60-inch Telescope at the Palomar
Observatory as part of the Zwicky Transient Facility project, ZTF is supported
by the National Science Foundation under Grants No. AST-1440341 and AST-2034437
and a collaboration including current partners Caltech, IPAC, the Oskar
Klein Center at Stockholm University, the University of Maryland, University
of California, Berkeley, the University of Wisconsin at Milwaukee, University
of Warwick, Ruhr University, Cornell University, Northwestern University, and
Drexel University. Operations are conducted by COO, IPAC, and UW.

\end{acknowledgements}

   \bibliographystyle{aa}
   \bibliography{aa60589-26}

\onecolumn

\begin{appendix}
\section{Frequencies from the Fourier analysis of the TESS light curves}
\label{app1}

The frequencies, amplitudes, and phases reported in this work were determined using a nonlinear least-squares fit to the model,
\begin{equation}
S(t)= \sum_{i=1}^N A_i \sin \left( 2\pi \left( \nu_i t + \phi_i \right)  \right) + c,
\end{equation}
where $N$ denotes the number of sinusoidal components, $A_i$, $\nu_i$, and $\phi_i$
are the amplitude, frequency, and phase of the $i$-th component,
respectively, and $c$ is a constant offset.
The phases $\phi_i$ are given for $t = \mathrm{BJD} - 2458216\ \mathrm{d}$.

\begin{figure}[ht!]
    \centering
    \begin{minipage}[t]{0.48\textwidth}
        \centering
        \includegraphics[width=\linewidth]{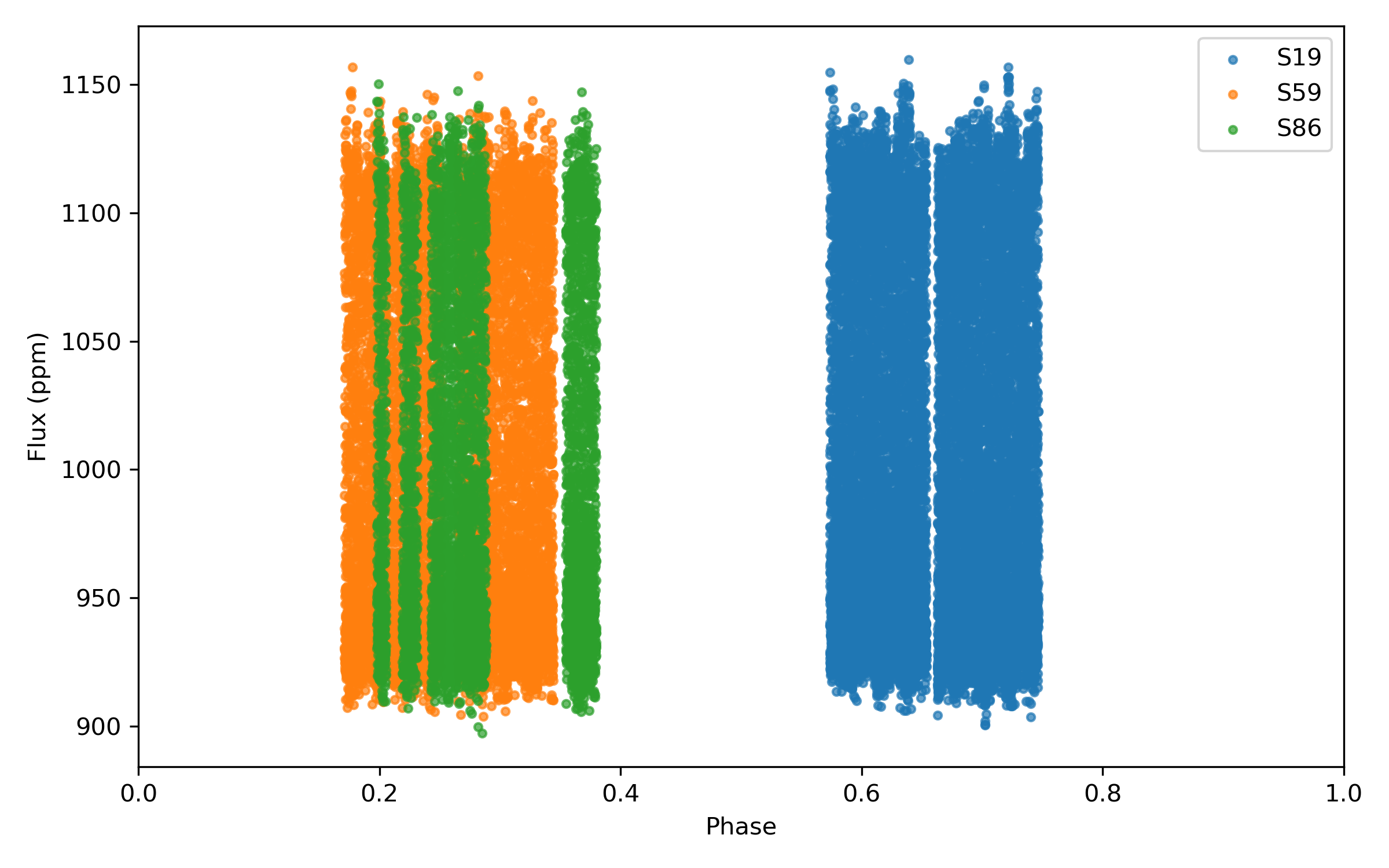}
        \caption{Orbital phase coverage of BL~Cam provided by the TESS observations in Sectors 19, 59, and 86. The large gaps between sectors result in incomplete sampling of the $\sim$144\,d orbital cycle.}
        \label{fig:orbital_phase_cover}
    \end{minipage}
    \hfill
    \begin{minipage}[t]{0.48\textwidth}
        \centering
        \includegraphics[width=\linewidth]{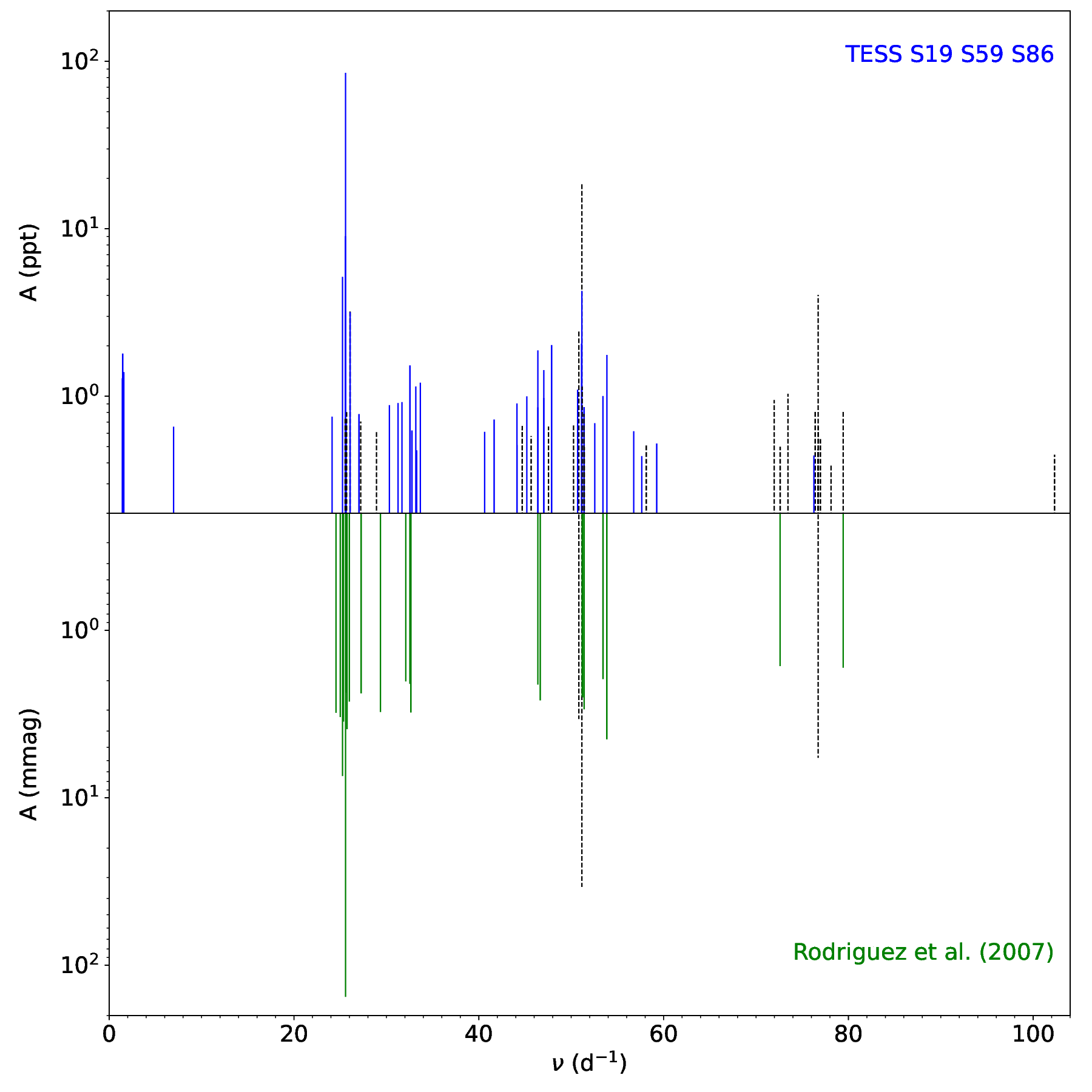}
        \caption{Comparison of the pulsation frequencies of BL~Cam detected in the ground-based photometry by \citet{2007A&A...471..255R} and in the combined TESS dataset (Sectors 19, 59, and 86). Independent frequencies are marked in blue and green, while black dashed lines indicate combination frequencies and harmonics.}
        \label{fig:full_TESS_vs_Rodriguez}
    \end{minipage}
\end{figure}

\begin{figure}[h!]
    \centering
    \includegraphics[width=0.32\textwidth]{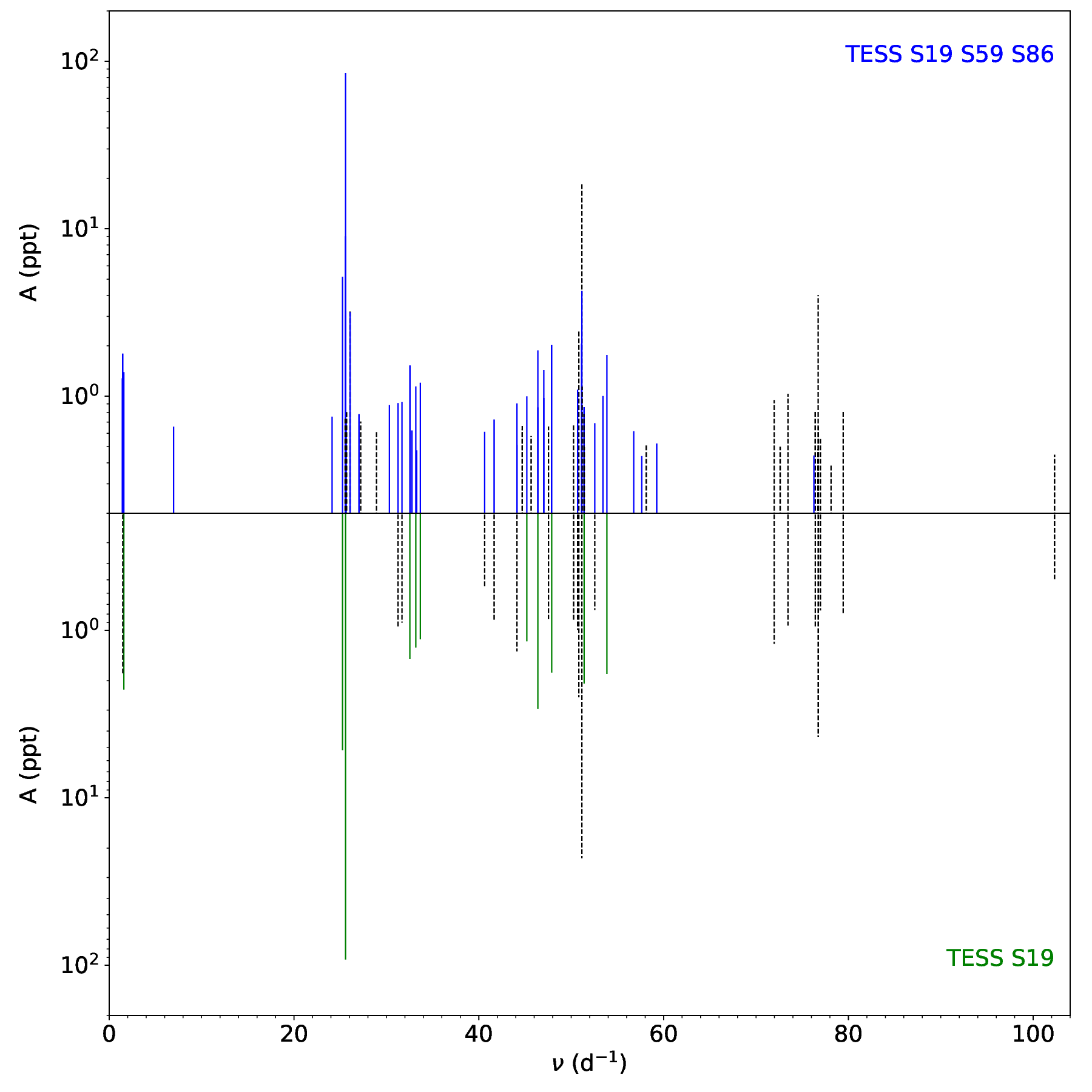}\hfill
    \includegraphics[width=0.32\textwidth]{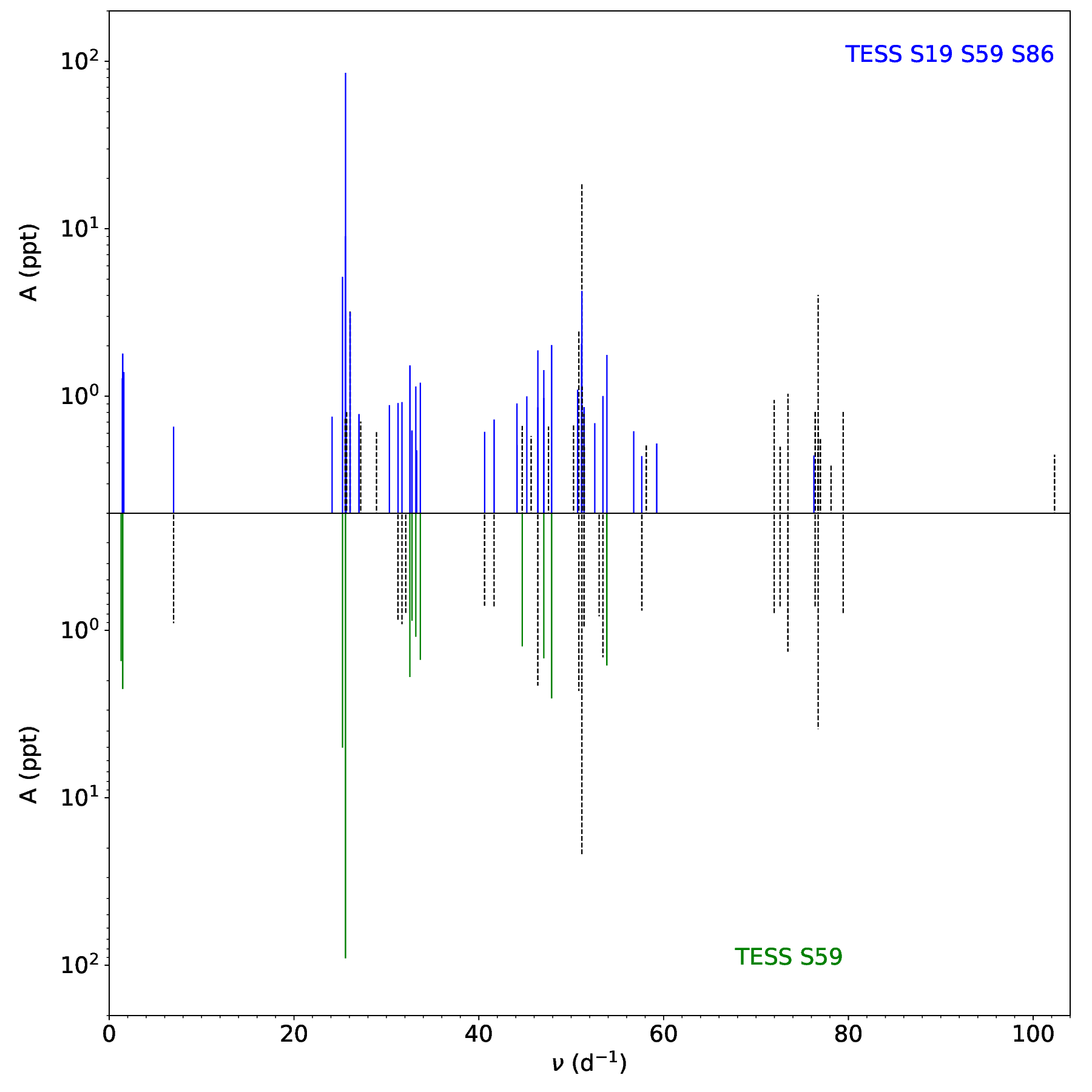}\hfill
    \includegraphics[width=0.32\textwidth]{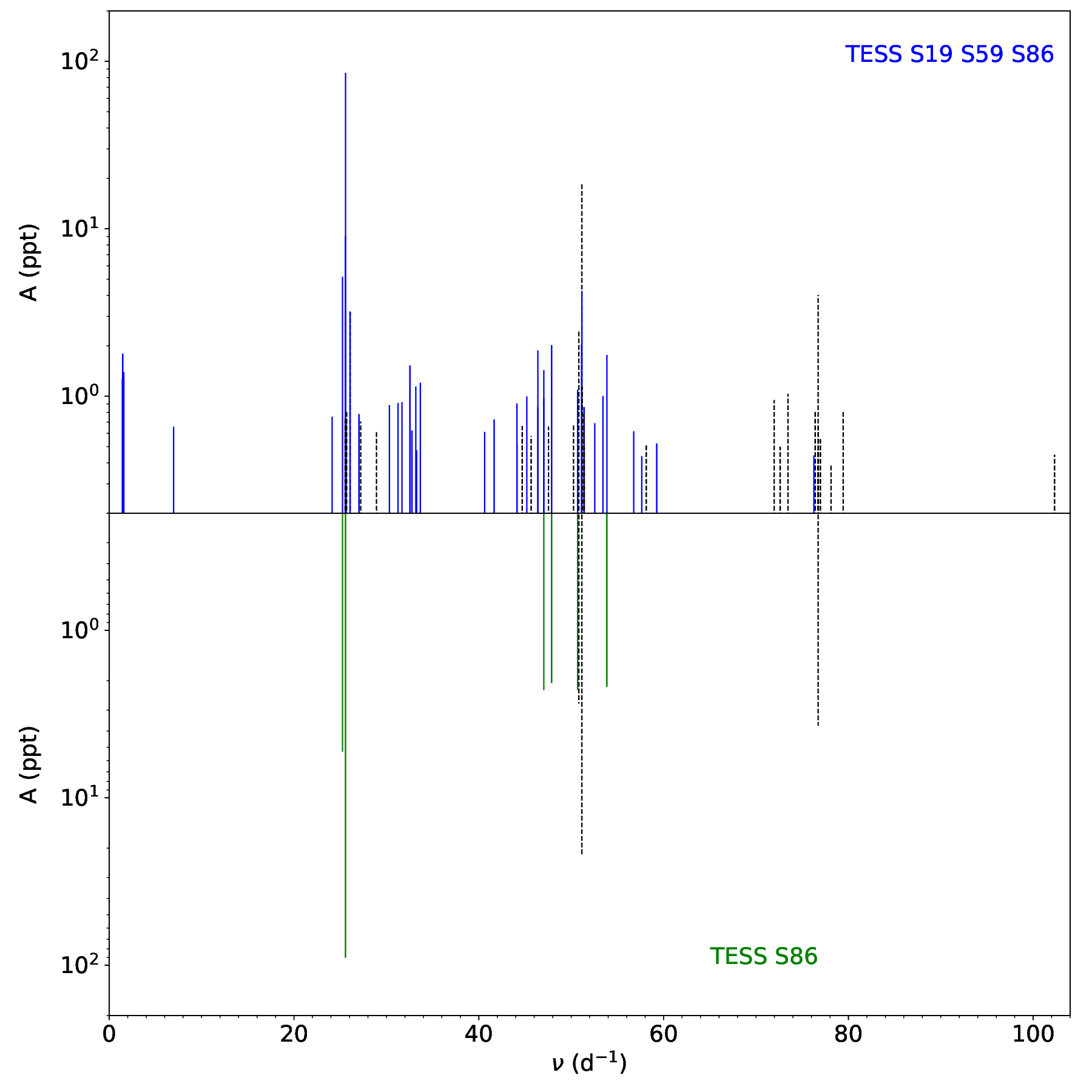}
    \caption{Comparison of the pulsation frequencies of BL~Cam obtained from the combined TESS dataset (Sectors 19, 59, and 86) and from individual sectors: Sector 19 (left panel),
    Sector 59 (middle panel), and Sector 86 (right panel).
    Independent frequencies are marked in blue and green, while black dashed lines indicate combination frequencies and harmonics.}
    \label{fig:full_TESSfull_vs_TESS_all}
\end{figure}

The orbital-phase coverage of BL~Cam provided by the TESS observations is shown in Fig.~\ref{fig:orbital_phase_cover}.
A comparison of the frequency sets obtained from different datasets is presented in Figs.\,\ref{fig:full_TESS_vs_Rodriguez} and \ref{fig:full_TESSfull_vs_TESS_all}.
In Fig.\,\ref{fig:full_TESS_vs_Rodriguez} we compare the frequencies derived from the combined TESS dataset (Sectors 19, 59, and 86)
with those reported from the ground-based photometry by \citet{2007A&A...471..255R}.
In Fig.\,\ref{fig:full_TESSfull_vs_TESS_all} we compare the combined TESS dataset with the frequencies
obtained from Sectors 19, 59, and 86 analysed individually.
For completeness, Table\,\ref{tab:freq_TESS_combined} lists the full frequency solution obtained from the combined TESS dataset,
including the fitted frequencies, amplitudes, and phases. We also provide the corresponding S/Ns
and indicate harmonic and combination terms where applicable.

\setlength{\LTcapwidth}{\textwidth}
\begin{longtable}{ccccccccc}
\caption{\label{tab:freq_TESS_combined}Frequency analysis results for BL~Cam from TESS photometry combining Sectors 19, 59, and 86.}
\\
\hline
ID & $\nu$ (d$^{-1}$) & $\sigma_{\nu}$ (d$^{-1}$) & $A$ (ppt) & $\sigma_A$ (ppt) & $\phi$ (0--1) & $\sigma_{\phi}$ & S/N & Remarks\\
\hline
\endfirsthead
\hline
ID & $\nu$ (d$^{-1}$) & $\sigma_{\nu}$ (d$^{-1}$) & $A$ (ppt) & $\sigma_A$ (ppt) & $\phi$ (0--1) & $\sigma_{\phi}$ & S/N & Remarks\\
\hline
\endhead
\hline
\endfoot
\multicolumn{9}{l}{The phases ($\phi$) are given for $t = \mathrm{BJD} - 2458216$\,d.}
\endlastfoot
\hline
  1 & 25.5768313 & 0.0000010 &  84.56 &  0.39 & 0.5842 & 0.0007 &  83.6 &  \\
  2 & 51.1536703 & 0.0000032 &  18.80 &  0.14 & 0.9702 & 0.0038 &  64.1 & $2\nu_{1}$ \\
  3 & 25.2521672 & 0.0000045 &   5.12 &  0.07 & 0.3730 & 0.0041 &  25.8 &  \\
  4 & 76.7304990 & 0.0000082 &   4.03 &  0.07 & 0.4364 & 0.0080 &  32.9 & $3\nu_{1}$ \\
  5 & 25.5958278 & 0.0000250 &   3.21 &  0.11 & 0.7745 & 0.0246 &  21.1 &  \\
  6 & 25.5669824 & 0.0000104 &   8.95 &  0.50 & 0.4428 & 0.0086 &  18.5 &  \\
  7 & 50.8290083 & 0.0000093 &   2.47 &  0.07 & 0.7954 & 0.0086 &  14.7 & $\nu_{1}+\nu_{3}$ \\
  8 & 25.5898902 & 0.0000205 &   3.03 &  0.15 & 0.5534 & 0.0212 &  17.3 &  \\
  9 & 46.3903196 & 0.0000314 &   1.86 &  0.11 & 0.1453 & 0.0459 &  14.9 &  \\
 10 & 47.8807799 & 0.0000112 &   2.00 &  0.06 & 0.9554 & 0.0104 &  12.9 &  \\
 11 & 51.1193450 & 0.0000119 &   2.04 &  0.07 & 0.5303 & 0.0110 &  13.2 &  \\
 12 & 53.8582150 & 0.0000128 &   1.75 &  0.06 & 0.1159 & 0.0119 &  12.8 &  \\
 13 &  1.4628200 & 0.0000128 &   1.78 &  0.07 & 0.5782 & 0.0118 &   7.6 &  \\
 14 & 51.1519880 & 0.0000145 &   4.22 &  0.14 & 0.9574 & 0.0169 &  11.8 &  \\
 15 & 32.5467821 & 0.0000149 &   1.51 &  0.07 & 0.3765 & 0.0138 &  13.1 &  \\
 16 &  1.5758062 & 0.0000164 &   1.38 &  0.07 & 0.7912 & 0.0151 &   7.0 &  \\
 17 & 51.3947754 & 0.0000567 &   0.85 &  0.09 & 0.1726 & 0.0724 &  10.8 &  \\
 18 & 33.6717031 & 0.0000188 &   1.19 &  0.06 & 0.7874 & 0.0175 &  11.8 &  \\
 19 &  1.4178523 & 0.0000181 &   1.27 &  0.07 & 0.0133 & 0.0167 &   5.6 &  \\
 20 & 33.1818652 & 0.0000200 &   1.13 &  0.06 & 0.7778 & 0.0185 &  11.1 &  \\
 21 & 50.6946114 & 0.0000210 &   1.08 &  0.07 & 0.4703 & 0.0195 &   8.9 &  \\
 22 & 47.0389501 & 0.0000673 &   0.97 &  0.23 & 0.8012 & 0.0656 &   8.5 &  \\
 23 & 73.4576208 & 0.0000217 &   1.03 &  0.06 & 0.4325 & 0.0202 &  10.3 & $\nu_{1}+\nu_{10}$ \\
 24 & 53.4412469 & 0.0000226 &   0.99 &  0.06 & 0.6696 & 0.0210 &   8.8 &  \\
 25 & 45.1983269 & 0.0000227 &   0.99 &  0.06 & 0.5882 & 0.0211 &   8.8 &  \\
 26 & 71.9671910 & 0.0000236 &   0.95 &  0.06 & 0.5510 & 0.0220 &  10.1 & $\nu_{1}+\nu_{9}$ \\
 27 & 51.1852611 & 0.0000214 &   1.14 &  0.07 & 0.5720 & 0.0198 &   8.0 & $\nu_{5}+\nu_{8}$ \\
 28 & 31.6826391 & 0.0000246 &   0.91 &  0.06 & 0.0749 & 0.0228 &   8.1 &  \\
 29 & 44.1338401 & 0.0000624 &   0.90 &  0.07 & 0.3058 & 0.0460 &   8.7 &  \\
 30 & 31.2559493 & 0.0000249 &   0.90 &  0.06 & 0.0241 & 0.0231 &   7.8 &  \\
 31 & 30.3213140 & 0.0000256 &   0.88 &  0.06 & 0.7422 & 0.0238 &   6.9 &  \\
 32 & 76.4058646 & 0.0000276 &   0.83 &  0.07 & 0.1764 & 0.0256 &   9.0 & $2\nu_{1}+\nu_{3}$ \\
 33 & 79.4350599 & 0.0000272 &   0.83 &  0.06 & 0.5500 & 0.0252 &   9.6 & $\nu_{1}+\nu_{12}$ \\
 34 & 25.6388047 & 0.0000432 &   0.70 &  0.07 & 0.9894 & 0.0383 &   5.7 &  \\
 35 & 25.6872955 & 0.0000315 &   0.80 &  0.07 & 0.2702 & 0.0280 &   6.1 & $-\nu_{8}+2\nu_{34}$ \\
 36 & 27.0324556 & 0.0000290 &   0.78 &  0.06 & 0.2876 & 0.0269 &   5.6 &  \\
 37 & 24.1176464 & 0.0000300 &   0.75 &  0.06 & 0.0743 & 0.0279 &   5.4 &  \\
 38 & 41.6570295 & 0.0000312 &   0.72 &  0.06 & 0.6102 & 0.0290 &   8.1 &  \\
 39 & 27.2183647 & 0.0000318 &   0.71 &  0.06 & 0.7736 & 0.0296 &   5.3 & $\nu_{13}+\nu_{17}-\nu_{34}$ \\
 40 & 25.5515303 & 0.0000389 &   2.17 &  0.32 & 0.4183 & 0.0392 &   5.2 &  \\
 41 & 52.5483682 & 0.0000328 &   0.68 &  0.06 & 0.7371 & 0.0305 &   6.3 &  \\
 42 & 51.4191266 & 0.0000524 &   0.50 &  0.07 & 0.3953 & 0.0529 &   6.1 & $-\nu_{3}+\nu_{11}+\nu_{40}$ \\
 43 & 47.5438323 & 0.0000341 &   0.66 &  0.06 & 0.5870 & 0.0317 &   6.1 & $\nu_{9}-\nu_{17}+\nu_{41}$ \\
 44 & 50.2553799 & 0.0000336 &   0.67 &  0.06 & 0.7711 & 0.0312 &   6.1 & $-2\nu_{6}+2\nu_{21}$ \\
 45 & 26.0678848 & 0.0000484 &   3.23 &  0.49 & 0.6044 & 0.0493 &   5.1 & $-3\nu_{19}+\nu_{31}$ \\
 46 & 47.0364018 & 0.0000459 &   1.42 &  0.23 & 0.8543 & 0.0447 &   6.3 &  \\
 47 &  6.9671378 & 0.0000344 &   0.65 &  0.06 & 0.7964 & 0.0320 &   5.9 &  \\
 48 & 46.3949570 & 0.0000687 &   0.85 &  0.11 & 0.3537 & 0.1005 &   6.5 &  \\
 49 & 76.7473294 & 0.0000458 &   0.66 &  0.07 & 0.0340 & 0.0441 &   7.1 & $\nu_{5}+\nu_{14}$ \\
 50 & 32.7626733 & 0.0000363 &   0.62 &  0.07 & 0.8923 & 0.0338 &   6.8 &  \\
 51 & 25.5336325 & 0.0000670 &   0.75 &  0.14 & 0.5494 & 0.0664 &   6.2 & $\nu_{1}+\nu_{5}-\nu_{34}$ \\
 52 & 26.0706239 & 0.0000497 &   3.15 &  0.49 & 0.4243 & 0.0506 &   5.0 &  \\
 53 & 28.9271594 & 0.0000361 &   0.62 &  0.06 & 0.0637 & 0.0336 &   5.1 & $\nu_{1}+\nu_{18}-\nu_{31}$ \\
 54 & 40.6296408 & 0.0000369 &   0.61 &  0.06 & 0.5029 & 0.0344 &   7.5 &  \\
 55 & 44.7004985 & 0.0000829 &   0.66 &  0.11 & 0.1563 & 0.0735 &   6.7 & $\nu_{22}-\nu_{24}+2\nu_{40}$ \\
 56 & 51.3857984 & 0.0000568 &   0.78 &  0.09 & 0.7245 & 0.0728 &   6.4 & $\nu_{8}-\nu_{47}+\nu_{50}$ \\
 57 & 56.7668970 & 0.0000367 &   0.61 &  0.06 & 0.3483 & 0.0341 &   6.6 &  \\
 58 & 45.6624291 & 0.0000389 &   0.58 &  0.06 & 0.2284 & 0.0361 &   6.1 & $-\nu_{11}+\nu_{30}+2\nu_{50}$ \\
 59 & 44.7019635 & 0.0000870 &   0.63 &  0.11 & 0.7410 & 0.0770 &   6.4 & $\nu_{6}-\nu_{9}+2\nu_{50}$ \\
 60 & 76.9643502 & 0.0000410 &   0.55 &  0.07 & 0.0878 & 0.0379 &   6.2 & $-\nu_{13}+\nu_{17}+\nu_{36}$ \\
 61 & 58.1218806 & 0.0000434 &   0.52 &  0.06 & 0.7824 & 0.0403 &   6.3 & $-\nu_{1}+\nu_{14}+\nu_{15}$ \\
 62 & 59.2541123 & 0.0000434 &   0.52 &  0.06 & 0.7108 & 0.0403 &   6.5 &  \\
 63 & 72.6160488 & 0.0000449 &   0.50 &  0.06 & 0.5856 & 0.0418 &   5.9 & $\nu_{1}+\nu_{22}$ \\
 64 & 44.1269258 & 0.0001105 &   0.51 &  0.07 & 0.6120 & 0.0814 &   5.7 &  \\
 65 & 33.2783231 & 0.0000481 &   0.47 &  0.06 & 0.1509 & 0.0445 &   5.6 &  \\
 66 & 76.7055189 & 0.0000545 &   0.48 &  0.07 & 0.2391 & 0.0534 &   5.5 & $2\nu_{1}+\nu_{40}$ \\
 67 & 102.3101505 & 0.0000502 &   0.45 &  0.06 & 0.2060 & 0.0467 &   6.0 & $\nu_{11}+\nu_{34}+\nu_{40}$ \\
 68 & 76.2686708 & 0.0000518 &   0.44 &  0.07 & 0.5946 & 0.0481 &   5.2 &  \\
 69 & 57.6433894 & 0.0000516 &   0.43 &  0.06 & 0.2177 & 0.0480 &   5.2 &  \\
 70 & 78.1251989 & 0.0000568 &   0.39 &  0.06 & 0.1929 & 0.0528 &   5.0 & $\nu_{1}+\nu_{41}$ \\
 \hline
\end{longtable}

\clearpage

\section{Asteroseismic modelling with Monte Carlo-based Bayesian Analysis}
\label{app2}

\begin{figure}[ht!]
	\includegraphics[width=\textwidth,clip]{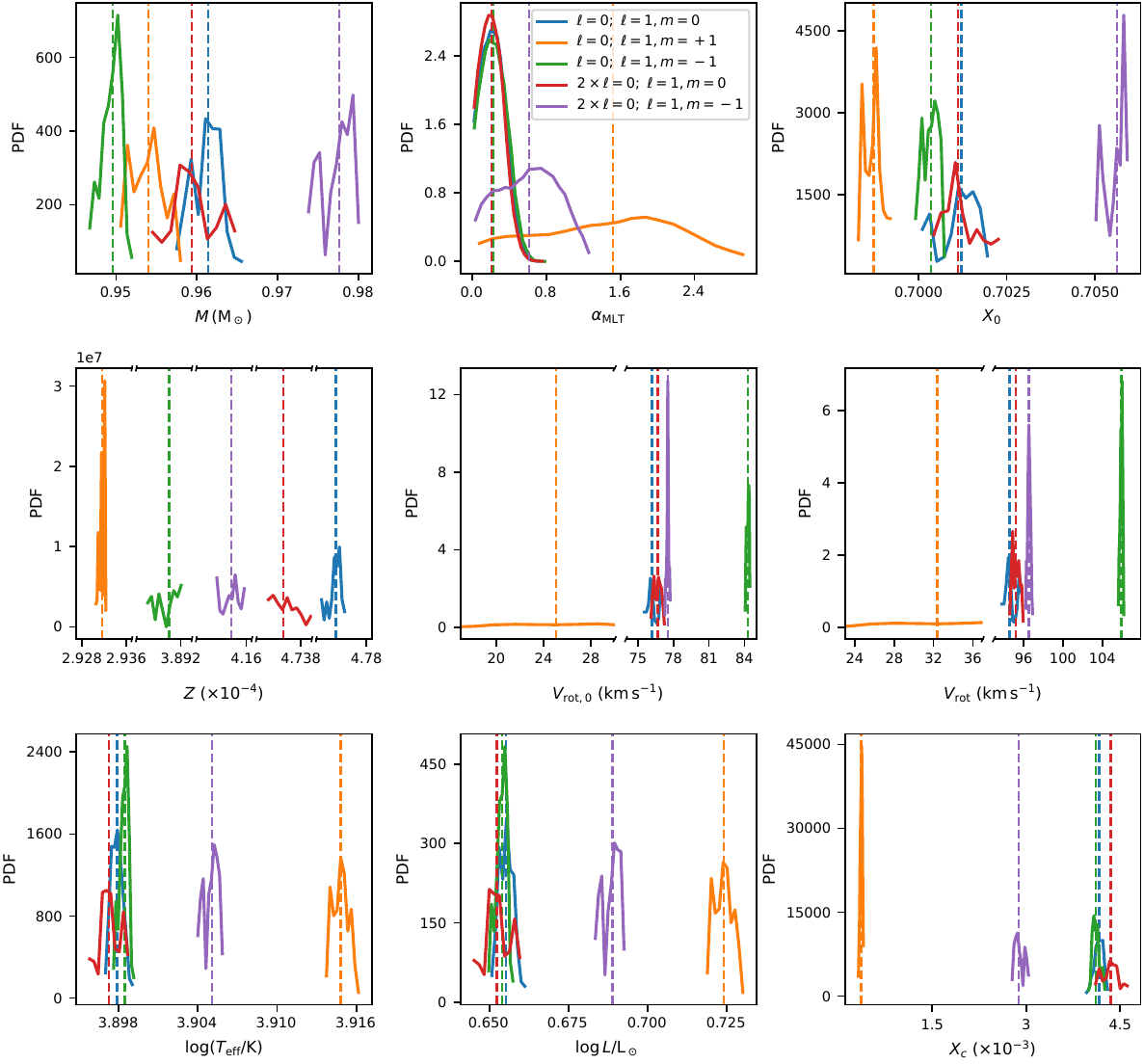}
	\caption{Posterior probability density functions (PDF) for the stellar parameters inferred from the seismic modelling of BL~Cam.
	The panels show (from top left to bottom right): mass $M$, mixing length parameter $\alpha_\mathrm{MLT}$, initial hydrogen abundance $X_0$,
	metallicity $Z$, initial equatorial velocity $V_\mathrm{rot, 0}$, present-day equatorial rotational velocity $V_\mathrm{rot}$,
	effective temperature $\log T_\mathrm{eff}$, luminosity $\log L/\mathrm{L}_{\sun}$, and
	central hydrogen abundance $X_c$.
	Different colours correspond to the five mode-identification scenarios.
	Vertical dashed lines indicate the median values.}
		\label{fig:hist_B1}
\end{figure}

\begin{figure}
	\includegraphics[width=\textwidth,clip]{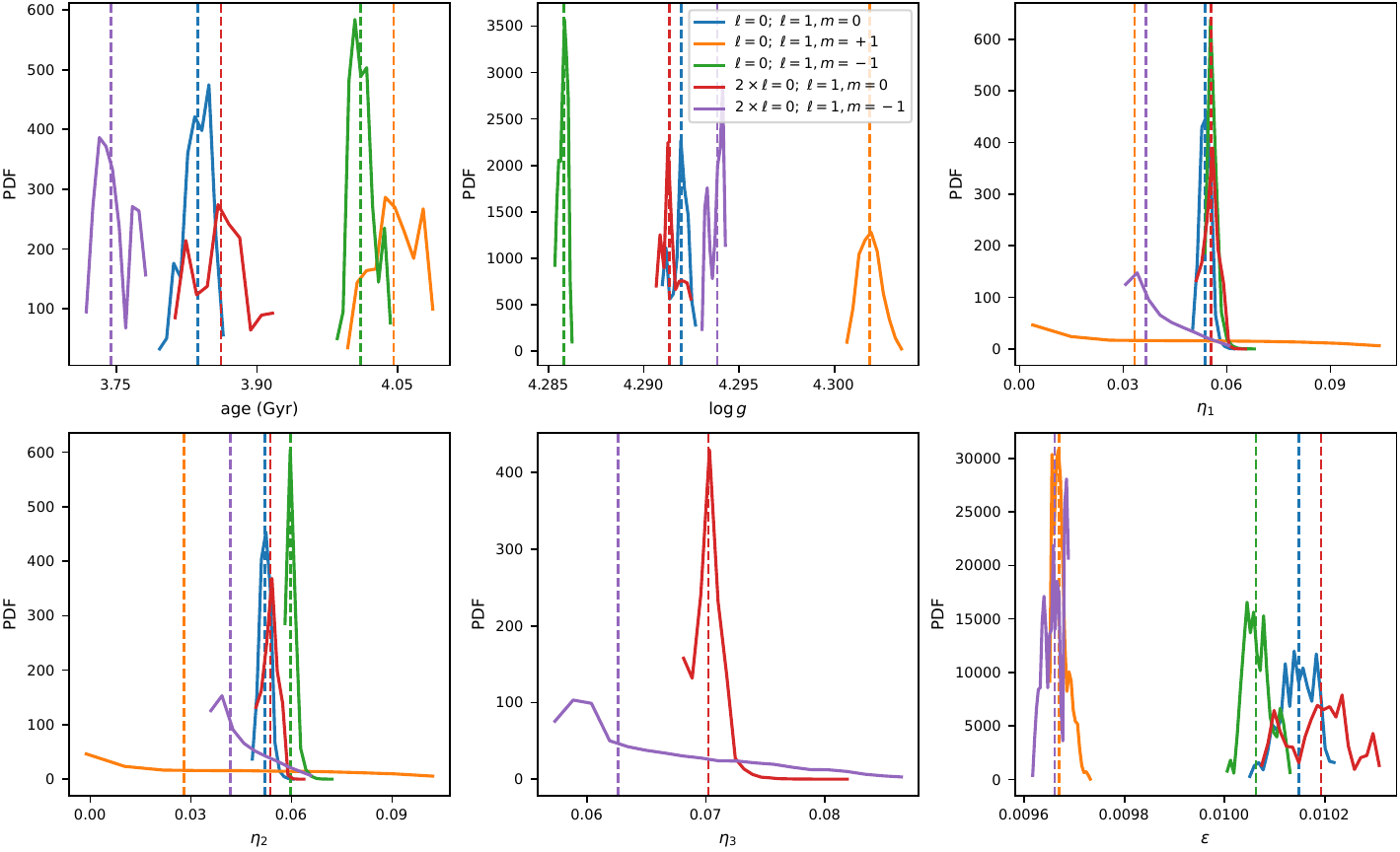}
	\caption{Same as in Fig.\,\ref{fig:hist_B1}, but for age, gravity $\log g$, the instability parameter  $\eta_1$ for mode $\nu_1$,
	instability parameter $\eta_2$ for mode $\nu_2$, $\eta_3$ for $\nu_8$, and the intrinsic mode amplitude $\varepsilon$ for $\nu_1$.}
		\label{fig:hist_B2}
\end{figure}

\end{appendix}

\end{document}